\documentclass[aps,pra,twocolumn,nobibnotes,superscriptaddress]{revtex4}
\usepackage[utf8]{inputenc}
\usepackage[T1]{fontenc}
\usepackage{bm}
\usepackage{ulem}
\usepackage{epsfig}
\usepackage{graphicx}
\usepackage{amssymb,amsmath,amsbsy,amsgen,amsfonts}
\usepackage{dcolumn}
\usepackage{amsthm}
\usepackage{mathtools}
\usepackage{mathrsfs}
\usepackage{latexsym}
\usepackage{array}
\usepackage{amstext}
\usepackage{multirow}
\usepackage{bbold}
\usepackage{color}
\usepackage{xcolor}
\usepackage{listings}
\usepackage{lipsum}
\usepackage{textcomp}
\usepackage{tikz}
\usetikzlibrary{shapes.geometric, arrows.meta, positioning}

\makeatletter
\allowdisplaybreaks[1]

\makeatletter
\let\old@makecaption=\@makecaption
\usepackage{caption}
\usepackage{subcaption}
\let\@makecaption=\old@makecaption
\makeatother

\usepackage{epstopdf} 

\DeclareSymbolFont{symbols}{OMS}{cmsy}{m}{n}

\newcommand{\secref}[1]{Sec.~\ref{#1}}
\newcommand{\appendref}[1]{Appendix~\ref{#1}}
\newcommand{\equatref}[1]{Eq.~(\ref{#1})}
\newcommand{\figref}[1]{Fig.~\ref{#1}}
\newcommand{\tabref}[1]{Table~\ref{#1}}

\definecolor{sciencered}{RGB}{180, 22, 44}
\definecolor{agrigreen}{RGB}{61, 138, 26}
\definecolor{engineeringyellow}{RGB}{255, 156, 47}
\definecolor{edublue}{RGB}{34, 61, 113}

\tikzstyle{smallblock} = [rectangle, rounded corners, minimum width=1cm, minimum height=0.55cm, text centered, draw=gray!60!black, thick, fill=gray!20]
\tikzstyle{medsmallblock} = [rectangle, rounded corners, minimum width=1.2cm, minimum height=0.55cm, text centered, draw=gray!60!black, thick, fill=gray!20]
\tikzstyle{mediumblock} = [rectangle, rounded corners, minimum width=1.8cm, minimum height=0.55cm, text centered, draw=gray!60!black, thick, fill=gray!20]
\tikzstyle{medlargeblock} = [rectangle, rounded corners, minimum width=1.8cm, minimum height=0.9cm, text width=1.6cm, text centered, draw=gray!60!black, thick, fill=gray!20]
\tikzstyle{largeblock} = [rectangle, rounded corners, minimum width=2.2cm, minimum height=0.9cm, text width=2cm, text centered, draw=gray!60!black, thick, fill=gray!20]

\begin{document}

\title{Demonstration of quantum random number generation using nitrogen vacancy centres}

\author{Conrad Strydom}
\email{conradstryd@gmail.com}
\affiliation{Stellenbosch Photonics Institute, Department of Physics, Stellenbosch University, Private Bag X1, Matieland 7602, South Africa}
\author{Mark Tame}
\affiliation{Stellenbosch Photonics Institute, Department of Physics, Stellenbosch University, Private Bag X1, Matieland 7602, South Africa}
\affiliation{National Institute for Theoretical and Computational Sciences (NITheCS) South Africa}

\begin{abstract}
Quantum random number generation~(QRNG) relies on the inherent unpredictability of quantum mechanical phenomena to efficiently generate high-quality random numbers that can be used in a wide range of cryptography and simulation applications.  Here we report the experimental demonstration of QRNG from the arrival times of photons emitted by nitrogen vacancy~(NV) centres in fluorescent nanodiamonds.  The generation rates achieved range from $0.173\,\text{Mbits/s}$ for a region with a single NV centre to $4.77\,\text{Mbits/s}$ for a region with just under 50 NV centres, where the latter demonstrates an order of magnitude improvement compared to the highest generation rate previously achieved with NV centres.  For all the regions investigated, the generated bits passed the ENT and NIST Statistical Test Suites without post-processing.  The results are consistent with our theoretical analysis, where we show that the min-entropy is very close to the ideal value of one per bit for all the regions investigated.  This work opens up new possibilities for robust QRNG in highly compact on-chip settings.
\end{abstract}


\maketitle

\section{Introduction}\label{sec:introduction} 

Random numbers are ubiquitous in cryptography~\cite{cryptography} and used extensively in Monte Carlo simulations~\cite{simulation}.  On conventional classical computers, random numbers are typically generated using pseudorandom number generators --- algorithms which use sophisticated deterministic formulas to produce seemingly random numbers by expanding an initial random seed~\cite{PRNG1, PRNG2, PRNG3}.  While pseudorandom number generators may be convenient to use, as they can be implemented in software on any classical computing platform, their fundamentally deterministic nature can result in the presence of underlying patterns in their output, making them more predictable than anticipated and compromising their utility.  For instance, given sufficient output from a standard pseudorandom number generator, modern machine learning algorithms are able to predict the future output of the generator with a high degree of accuracy, making standard pseudorandom number generators unsuitable for cryptographic applications~\cite{PRNGpredict1, PRNGpredict2}.  Cryptographically secure pseudorandom number generators can be specifically designed to avoid predictability, but they are resource intensive and are therefore often implemented on dedicated hardware, such as field-programmable gate arrays (FPGAs)~\cite{CSPRNG1, CSPRNG2}.  Furthermore, a recent study suggests that even cryptographically secure pseudorandom number generators may still be at risk of becoming compromised as machine learning algorithms continue to improve~\cite{CSPRNGpredict}.

In contrast, quantum random number generation~(QRNG) relies on the inherent unpredictability of quantum mechanical phenomena to generate random numbers which are truly unpredictable~\cite{QRNGreview1, QRNGreview2, QRNGreview3}.  The first QRNG systems that were developed during the second half of the twentieth century relied on the spontaneous decay of radioactive isotopes~\cite{QRNGsd1, QRNGsd2, QRNGsd3}.  While these early systems were capable of generating high-quality random numbers, they were generally quite bulky and their generation rates were limited by the decay rates of the isotopes used.  In recent years, there has been a growing interest in implementing QRNG schemes on cloud-based quantum computers, including superconducting systems~\cite{superconducting1, superconducting2, superconducting3} and trapped ions~\cite{trappedions}.  Although cloud-based quantum computers offer a sophisticated and easily-accessible platform for QRNG, these systems do not provide the ability to generate private random keys locally, which is essential in cryptography.  Thus the development of dedicated locally-available quantum hardware for QRNG is highly important for cryptographic applications.

Photonic systems have emerged as the standard hardware platform for QRNG.  Various photonic QRNG schemes have been developed and successfully demonstrated experimentally.  These include schemes based on single-photon phenomena, such as the branching paths scheme~\cite{QRNGbp1, QRNGbp2, QRNGbp3, QRNGbp4} and the time-of-arrival scheme~\cite{QRNGtoa1, QRNGtoa2, QRNGtoa3, QRNGtoa4, QRNGtoa5, QRNGtoa6, QRNGtoa7, QRNGtoa8}, as well as schemes based on quantum measurements of continuous variables, such as the vacuum fluctuation scheme~\cite{QRNGvf1, QRNGvf2, QRNGvf3, QRNGvf4} and the laser phase fluctuation scheme~\cite{QRNGlpf1, QRNGlpf2, QRNGlpf3, QRNGlpf4}.  More recently self-testing schemes~\cite{QRNGst1, QRNGst2, QRNGst3, QRNGst4}, in which a Bell test is employed to certify the randomness of the output, have also become popular.  In this work, we experimentally demonstrate the time-of-arrival scheme for QRNG with photons emitted by nitrogen vacancy~(NV) centres in fluorescent nanodiamonds.  While the time-of-arrival scheme offers neither the high generation rates of continuous-variable schemes nor the verifiable security of self-testing schemes, it remains attractive as a source of high-quality randomness due to its simplicity and robustness.

NV centres~\cite{NVCreview1, NVCreview2} are defect centres in the carbon lattice structure of diamond, where a substitutional nitrogen atom and a missing carbon atom or vacancy is present in the place of two adjacent carbon atoms.  NV centres can be either neutral, which is often denoted by $\text{NV}^0$, or negatively charged, which is often denoted by $\text{NV}^-$.  Both $\text{NV}^0$ and $\text{NV}^-$ centres possess energy levels, similar to those of atoms and ions~\cite{NVClevels}, which has led to them being dubbed `artificial' atoms.  Their energy level structure is what enables them to serve as single-photon sources.  In particular, when an isolated NV centre has been driven into its excited state by an external driving field (usually in the form of a pulsed or continuous-wave (CW) laser) it spontaneously decays back to its ground state and emits a single photon.  Single photons emitted by NV centres have been successfully employed in quantum key distribution~\cite{NVCQKD1, NVCQKD2} applications and probing fundamental concepts in quantum plasmonics~\cite{NVCplasmonics1, NVCplasmonics2} and most recently entanglement generation~\cite{NVCentanglement}.  The spin states of $\text{NV}^-$ centres have also been utilised in quantum memories~\cite{NVCmemory1, NVCmemory2}, as well as in a wide variety of quantum sensing applications~\cite{NVCsensing}, such as magnetometry~\cite{NVCmagnetometry1, NVCmagnetometry2, NVCmagnetometry3} and thermometry~\cite{NVCthermometry}.  They have been proposed for many other novel applications, including dark matter detection~\cite{NVCdarkmatter} and quantum gravity tests~\cite{NVCgravity1, NVCgravity2}.

Recently, NV centres and other single-photon sources have been used with the branching paths scheme for QRNG using the photons emitted~\cite{QRNGbpNVC1, QRNGbpNVC2, QRNGbpSPE1, QRNGbpSPE2}.  The branching paths scheme has also been modelled and analysed theoretically for a single-photon source~\cite{QRNGbpSPEtheory}.  However, an alternative approach is the time-of-arrival scheme~\cite{QRNGtoa1, QRNGtoa2, QRNGtoa3, QRNGtoa4, QRNGtoa5, QRNGtoa6, QRNGtoa7, QRNGtoa8}, which differs significantly from the branching paths approach and has three major advantages.  Firstly, the time-of-arrival scheme can be implemented with arguably the simplest optical setup --- it requires only a source and a single-photon detector.  Secondly, much higher random number generation rates can be achieved with the time-of-arrival scheme than with the branching paths scheme, as multiple bits can be extracted from each random photon's arrival time.  Thirdly, the time-of-arrival scheme is one of the few schemes that can produce high-quality uniform random numbers without randomness extraction or any other form of post-processing.  The time-of-arrival scheme has not yet been realised experimentally with single photons emitted by NV centres.  Previously it has been implemented with photons from a highly attenuated coherent source.  As such, the time-of-arrival scheme is only well understood for a coherent source, despite being based on single-photon phenomena.

In this paper, we report on an experimental demonstration of the time-of-arrival scheme with photons emitted by NV centres and present a theoretical analysis of QRNG based on the arrival times of photons from single-photon emitters.  In our experiments, we employed a laser-scanning confocal microscopy setup to excite NV centres in fluorescent nanodiamonds and collect the emitted photons.  To explore the possibility of increasing the random number generation rate at the cost of multi-photon contamination, we investigated not only single photons emitted by an isolated NV centre, but also photons collected from regions with multiple NV centres.  The generation rates achieved range from $0.173\,\text{Mbits/s}$ for a single NV centre to $4.77\,\text{Mbits/s}$ for a cluster consisting of just under 50 NV centres.  For all the regions considered, the generated bits were able to pass industry standard tests without any form of post-processing.  This is in excellent agreement with our theoretical analysis, where we show that, under the experimental conditions, the min-entropy is very close to the ideal value of one per bit for all the regions considered.  While our experiments are limited to NV centres, our experimental and theoretical results extend naturally to other defect centres in diamond and many other single-photon sources.  Our study will be of interest to researchers investigating robust, moderate speed QRNG in highly compact on-chip settings and to researchers exploring the practical utility of solid-state single-photon sources, such as NV centres in diamond, both experimentally and theoretically.

The paper is structured as follows.  In \secref{sec:setup}, we describe the experimental setup used to probe NV centres in fluorescent nanodiamonds.  In \secref{sec:characterisation}, we present a basic characterisation of the different regions considered in our implementation of the time-of-arrival scheme with photons emitted by NV centres.  The results obtained for these different regions are presented in \secref{sec:QRNG}, along with our theoretical analysis of the time-of-arrival scheme for single-photon emitters.  Concluding remarks and possible future extensions of our work are discussed in \secref{sec:conclusion}.  Additional details about the regions considered in the experiments, as well as further experimental and theoretical results are given in the appendices.

\section{Experimental setup}\label{sec:setup} 

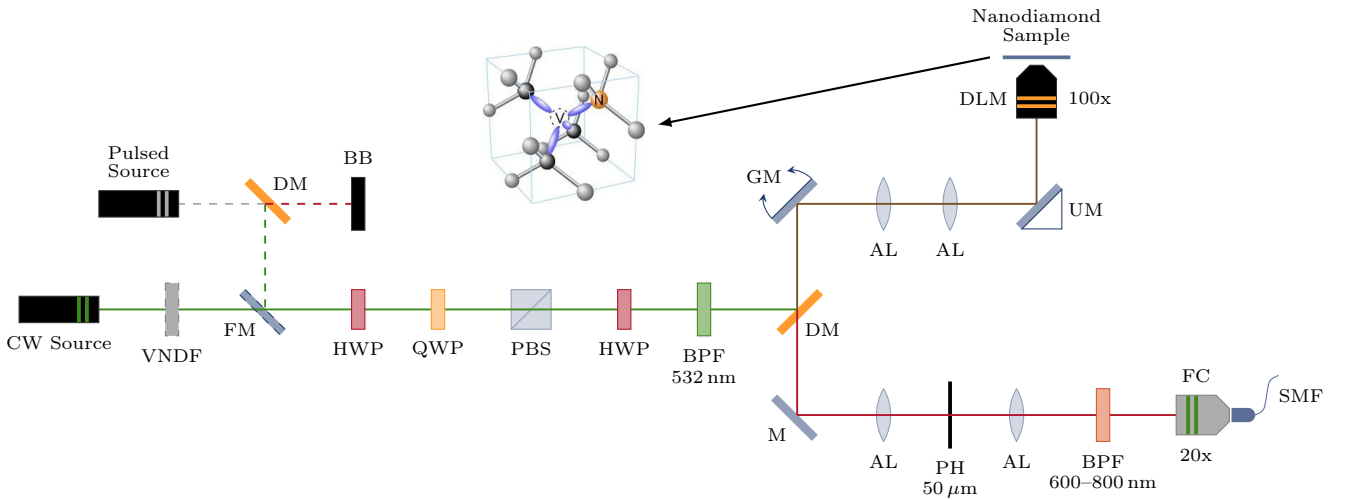
\begin{figure*}
    \centering
    \begin{tikzpicture}[scale=0.35, font=\scriptsize]
    \draw[edublue!50, fill=edublue!50] (7.5, 0) -- (9, -1.5) -- (8.8, -1.7) -- (7.3, -0.2) -- cycle;
    \draw[engineeringyellow, fill=engineeringyellow] (7.5, 4) -- (9, 2.5) -- (9.2, 2.7) -- (7.7, 4.2) -- cycle;
    \draw[edublue!50, fill=edublue!20] (17.5, 0) rectangle (19, -1.5);
    \draw[edublue!50] (17.5, -1.5) -- (19, 0);
    \draw[engineeringyellow, fill=engineeringyellow] (27.5, -1.5) -- (29, 0) -- (29.2, -0.2) -- (27.7, -1.7) -- cycle;
    \draw[edublue!50, fill=edublue!50] (27.5, 2.5) -- (29, 4) -- (28.8, 4.2) -- (27.3, 2.7) -- cycle;
    \draw[edublue!50, fill=edublue!20] (31.5, 4.25) arc (30:-30:2) arc (210:150:2);
    \draw[edublue!50, fill=edublue!20] (34, 4.25) arc (30:-30:2) arc (210:150:2);
    \draw[edublue!50, fill=edublue!50] (36.5, 2.5) -- (38, 4) -- (38.2, 3.8) -- (36.7, 2.3) -- cycle;
    \draw[edublue!50, fill=edublue!50] (27.5, -4) -- (29, -5.5) -- (28.8, -5.7) -- (27.3, -4.2) -- cycle;
    \draw[edublue!50, fill=edublue!20] (31.5, -3.75) arc (30:-30:2) arc (210:150:2);
    \draw[black, fill=black] (33.95, -3.5) rectangle (34.05, -6);
    \draw[edublue!50, fill=edublue!20] (36.5, -3.75) arc (30:-30:2) arc (210:150:2);
    \draw[agrigreen, line width=0.25mm] (2, -0.75) -- (28.25, -0.75);
    \draw[dashed, agrigreen, line width=0.25mm] (8.25, 3.25) -- (8.25, -0.75);
    \draw[dashed, gray!65, line width=0.25mm] (8.25, 3.25) -- (5, 3.25);
    \draw[dashed, sciencered, line width=0.25mm] (8.25, 3.25) -- (11.5, 3.25);
    \draw[agrigreen!50!sciencered, line width=0.25mm] (28.25, -0.75) -- (28.25, 3.25);
    \draw[agrigreen!50!sciencered, line width=0.25mm] (28.25, 3.25) -- (37.25, 3.25);
    \draw[agrigreen!50!sciencered, line width=0.25mm] (37.25, 3.25) -- (37.25, 6.5);
    \draw[sciencered, line width=0.25mm] (28.25, -0.75) -- (28.25, -4.75);
    \draw[sciencered, line width=0.25mm] (28.25, -4.75) -- (42.5, -4.75);
    \draw[black!80, fill=black] (-1, -0.25) rectangle (2, -1.25);
    \draw[agrigreen, fill=agrigreen] (1.2, -0.29) rectangle (1.3, -1.21);
    \draw[agrigreen, fill=agrigreen] (1.5, -0.29) rectangle (1.6, -1.21);
    \draw[dashed, gray, fill=gray!65] (4.5, 0.25) rectangle (5, -1.75);
    \draw[thin, dashed, edublue] (7.5, 0) -- (9, -1.5) -- (8.8, -1.7) -- (7.3, -0.2) -- cycle;
    \draw[black!80, fill=black] (2, 2.75) rectangle (5, 3.75);
    \draw[gray!65, fill=gray!65] (4.2, 3.71) rectangle (4.3, 2.79);
    \draw[gray!65, fill=gray!65] (4.5, 3.71) rectangle (4.6, 2.79);
    \draw[black!80, fill=black] (11.5, 2.25) rectangle (12, 4.25);
    \draw[sciencered, fill=sciencered!50] (11.5, 0) rectangle (12, -1.5);
    \draw[engineeringyellow, fill=engineeringyellow!50] (14.5, 0) rectangle (15, -1.5);
    \draw[sciencered, fill=sciencered!50] (21.5, 0) rectangle (22, -1.5);
    \draw[agrigreen, fill=agrigreen!50] (24.5, 0.25) rectangle (25, -1.75);
    \draw[thin, edublue] (28.8, 4.2) -- (27.3, 2.7);
    \draw[thin, -stealth, edublue] (28.8, 4.2) arc (45:100:1.06);
    \draw[thin, -stealth, edublue] (27.3, 2.7) arc (225:170:1.06);
    \draw[thin, edublue] (38.2, 3.8) -- (36.7, 2.3) -- (38.2, 2.3) -- cycle;
    \draw[black!80, fill=black] (36.5, 6.5) -- (38, 6.5) -- (38, 8) -- (37.5, 8.5) -- (37, 8.5) -- (36.5, 8) -- cycle;
    \draw[engineeringyellow, fill=engineeringyellow] (36.54, 6.9) rectangle (37.96, 7);
    \draw[engineeringyellow, fill=engineeringyellow] (36.54, 7.2) rectangle (37.96, 7.3);
    \draw[edublue!75, fill=edublue!75] (36, 8.75) rectangle (38.5, 8.85);
    \draw[engineeringyellow!50!sciencered, fill=engineeringyellow!50!sciencered!50] (39.5, -3.75) rectangle (40, -5.75);
    \draw[gray, fill=gray!65] (42.5, -4) -- (44, -4) -- (44.5, -4.5) -- (44.5, -5) -- (44, -5.5) -- (42.5, -5.5) -- cycle;
    \draw[agrigreen, fill=agrigreen] (42.9, -4.04) rectangle (43, -5.46);
    \draw[agrigreen, fill=agrigreen] (43.2, -4.04) rectangle (43.3, -5.46);
    \draw[edublue!75, fill=edublue!75] (44.6, -4.5) -- (44.6, -5) -- (45.2, -5) arc (-90:90:0.25) -- cycle;
    \draw[edublue!75] plot [smooth, tension=0.5] coordinates {(45.45, -4.75) (45.8, -4.5) (46, -3.5) (46.35, -3.25)};
    \node[] at (0.5, -2) {CW Source};
    \node[] at (4.75, -2.5) {VNDF};
    \node[] at (7.3, -1.5) {FM};
    \node[] at (9.2, 4) {DM};
    \node[] at (3.5, 5.2) {Pulsed};
    \node[] at (3.5, 4.5) {Source};
    \node[] at (11.75, 5) {BB};
    \node[] at (11.75, -2.25) {HWP};
    \node[] at (14.75, -2.25) {QWP};
    \node[] at (18.25, -2.25) {PBS};
    \node[] at (21.75, -2.25) {HWP};
    \node[] at (24.75, -2.5) {BPF};
    \node[] at (24.75, -3.3) {532$\,\text{nm}$};
    \node[] at (29.2, -1.5) {DM};
    \node[] at (27, 4.3) {GM};
    \node[] at (31.5, 1.5) {AL};
    \node[] at (34, 1.5) {AL};
    \node[] at (39.1, 3.05) {UM};
    \node[] at (35.25, 7.25) {DLM};
    \node[] at (39.25, 7.25) {100x};
    \node[] at (37.25, 10.4) {Nanodiamond};
    \node[] at (37.25, 9.6) {Sample};
    \node[] at (27.5, -5.5) {M};
    \node[] at (31.5, -6.5) {AL};
    \node[] at (34, -6.75) {PH};
    \node[] at (34, -7.6) {50$\,\mu\text{m}$};
    \node[] at (36.5, -6.5) {AL};
    \node[] at (39.75, -6.5) {BPF};
    \node[] at (39.75, -7.3) {600--800$\,\text{nm}$};
    \node[] at (43.25, -3.25) {FC};
    \node[] at (43.25, -6.25) {20x};
    \node[] at (47.2, -4) {SMF};
    \node[] at (19.5, 6.25) {\includegraphics[scale=0.4]{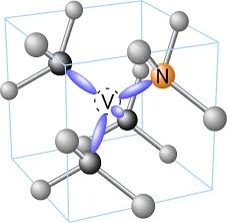}};
    \draw[-latex, thick] (35.5, 8.8) -- (23, 6.25);
    \end{tikzpicture}
    \caption{Laser-scanning confocal microscopy setup used to excite NV centres in fluorescent nanodiamonds and collect the emitted photons.  The following abbreviations are used: continuous-wave~(CW), variable neutral density filter~(VNDF), dichroic mirror~(DM), beam block~(BB), flip mirror~(FM), half-wave plate~(HWP), quarter-wave plate~(QWP), polarising beamsplitter~(PBS), bandpass filter~(BPF), galvonometric mirror~(GM), achromatic lens~(AL), upright mirror~(UM), diffraction-limited microscope~(DLM), mirror~(M), pinhole~(PH), fibre coupler~(FC) and single-mode fibre~(SMF).}
    \label{fig:setup}
\end{figure*}

The laser-scanning confocal microscopy setup~\cite{NVCsetupbook, NVCsetupthesis} which we use to excite NV centres in fluorescent nanodiamonds and collect the emitted photons is shown in \figref{fig:setup}.  In this setup, NV centres can be excited either with a CW source or with a pulsed source.  The CW source is a green CW laser (Thorlabs CPS532) with a central wavelength of $532\,\text{nm}$.  In all our experiments, we use a variable neutral density filter~(VNDF) to adjust the intensity of this laser so that a power of around $300\,\mu\text{W}$ is measured just before the upright mirror~(UM), at the entrance to the vertical microscope.  The pulsed source is a pulsed supercontinuum white laser (NKT Photonics Fianium WhiteLase Micro) with a pulse rate of $23.8\,\text{MHz}$ and a pulse duration of around $6\,\text{ps}$.  Light from the pulsed white laser with wavelengths shorter than $567\,\text{nm}$, which includes the wavelengths around $532\,\text{nm}$ that are needed to excite NV centres~\cite{NVCentanglement, QRNGbpNVC1}, is reflected into the main part of the setup by a dichroic mirror~(DM) (Thorlabs DMLP567R).  The redundant longer wavelengths (above $567\,\text{nm}$) are transmitted though the DM and absorbed by a beam block~(BB).  A flip mirror~(FM) enables us to switch between exciting NV centres with the CW source or the pulsed source.  Unless specifically stated otherwise, experiments in this paper were carried out with the CW source.

The excitation pump beam is sent through a half-wave plate~(HWP), a quarter-wave plate~(QWP), a polarising beamsplitter~(PBS), another HWP and an optical bandpass filter~(BPF).  The first three components clean up the polarisation of the pump beam.  In particular, the PBS acts as a horizontal polariser, while the preceding HWP and QWP are used to adjust the polarisation so as to minimise loss in the PBS.  The HWP after the PBS enables the tuning of the polarisation of the pump beam such that photon emission from the NV centres being excited is maximised~\cite{NVCpolarisation}.  The BPF (Thorlabs FL532-10) has a passband centred around $(532\pm10)\,\text{nm}$ and is used to remove any redundant wavelengths that will not meaningfully contribute to the excitation of NV centres.  While the BPF helps to purify both sources, it is most critical for the pulsed source, which at this point still contains a broad range of visible wavelengths below $567\,\text{nm}$.  Finally, another DM, a 2D galvonometric mirror~(GM) and a UM direct the pump beam into a vertical microscope, where a 100x diffraction-limited microscope~(DLM) objective is used to focus the pump beam onto a sample of nanodiamonds.  The DLM objective (Olympus UPlanFL~N) is an infinity-corrected, high numerical aperture ($\text{NA} = 1.3$), oil-immersion microscope objective.

Our sample of nanodiamonds was prepared by spin coating a sonicated suspension of nanodiamonds in deionised water (Ad\'{a}mas Nanotechnologies ND-NV-40nm-Low), diluted to a concentration of $0.2\,\text{mg/m}\ell$, onto a high-precision $\#1.5$~thickness glass coverslip (Thorlabs CG15XH1).  The nanodiamonds have an average diameter of around $40\,\text{nm}$ and each nanodiamond can contain between one and four NV centres, if it contains any.  Approximately $59\%$ of these NV centres are $\text{NV}^0$ centres and the remaining $41\%$ are $\text{NV}^-$ centres.  When the pump beam is focused onto a region of the sample with a nanodiamond or a cluster of nanodiamonds which contains one or more NV centres, it will drive transitions from the ground state to the excited state.  Each NV centre in the nanodiamond (or cluster) will emit a photon each time it spontaneously decays back to its ground state, causing the nanodiamond(s) to fluoresce~\cite{NVClevels}.

Light from the sample, which includes both fluorescence from the nanodiamonds and scattered light from the pump beam, is collected by the same DLM objective that was used to focus the pump beam onto the sample.  In particular, light from different points on the sample exit the rear aperture of the DLM objective as diverging collimated beams, thus forming an expanding image of the sample focused at infinity.  An arrangement of achromatic lenses~(ALs) is used to collect these diverging beams, so that light from the sample can be captured.  Since NV centres have a very broad fluorescence spectrum at room temperature~\cite{NVCreview1, NVClevels, NVCentanglement}, achromatic doublet lenses are needed to mitigate chromatic aberration.  This AL arrangement is discussed in Ref.~\cite{NVCsetupthesis}, and described in more detail in \appendref{append:confocal}.  The DM ensures that only fluorescence from the nanodiamonds, which consists mainly of wavelengths above $567\,\text{nm}$~\cite{NVCreview1, NVClevels, NVCentanglement}, is retained in the collection.  The $50\,\mu\text{m}$ pinhole~(PH), in combination with the 100x DLM objective and the ALs, enables us to pick out a circular capture region with a diameter of $500\,\text{nm}$ on the sample~\cite{NVCentanglement, NVCsetupbook}.  A final AL, placed one focal length from the PH, is used to collimate the light from the capture region that passes though the PH.

The collimated beam then passes though a broadband optical BPF, with a passband from $600\,\text{nm}$ to $800\,\text{nm}$.  The filter is designed to transmit only fluorescence from the nanodiamonds, which falls within this range~\cite{NVCreview1, NVClevels, NVCentanglement}, and to block any light from the pump beam which is able to leak through the DM, as well as other ambient light.  This custom BPF is a combination of three filters, namely a high-pass filter (Thorlabs FELH600) with a cutoff wavelength of $600\,\text{nm}$, a low-pass filter (Thorlabs FESH800) with a cutoff wavelength of $800\,\text{nm}$ and a bandstop filter (Thorlabs NF533-17) with a stopband centred around $(533\pm17)\,\text{nm}$.  The high-pass filter and the low-pass filter combine to give the desired passband.  The bandstop filter blocks any light from the pump beam which leaks through both the DM and the other filters.  Finally, the beam is reflected into a fibre coupler~(FC), which is connected to a single-mode optical fibre~(SMF).  This SMF can be connected in different configurations, depending on the analysis to be carried out.  Various types of analysis, needed to identify and characterise regions of interest on the nanodiamond sample, are discussed in the next section.

\section{Characterisation of regions}\label{sec:characterisation} 

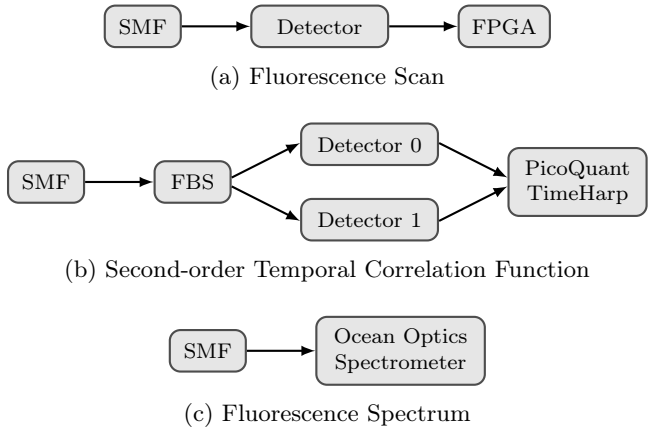
\begin{figure}
    \centering
    \begin{subfigure}[b]{.48\textwidth}
        \centering
        \begin{tikzpicture}[scale=0.35, font=\footnotesize]
        \node (SMF) [smallblock] {SMF};
        \node (Det) [mediumblock, right=0.9cm of SMF] {Detector};
        \node (FPGA) [medsmallblock, right=0.9cm of Det] {FPGA};
        \draw [-latex, thick] (SMF) -- (Det);
        \draw [-latex, thick] (Det) -- (FPGA);
        \end{tikzpicture}
        \caption{Fluorescence Scan}
        \label{fig:scanconfig}
    \end{subfigure}

    \vspace{0.3cm}
    
    \begin{subfigure}[b]{.48\textwidth}
        \centering
        \begin{tikzpicture}[scale=0.35, font=\footnotesize]
        \node (SMF) [smallblock] {SMF};
        \node (FBS) [smallblock, right=0.9cm of SMF] {FBS};
        \node (Det0) [mediumblock, right=0.9cm of FBS, yshift=0.5cm] {Detector 0};
        \node (Det1) [mediumblock, right=0.9cm of FBS, yshift=-0.5cm] {Detector 1};
        \node (TH) [medlargeblock, right=0.9cm of Det0, yshift=-0.5cm] {PicoQuant TimeHarp};
        \node (RefFBS0) [left=0.9cm of Det0, yshift=-0.5cm] {};
        \node (RefFBS1) [left=0.9cm of Det1, yshift=0.5cm] {};
        \node (RefDet0F) [right=0.9cm of FBS, yshift=0.575cm] {};
        \node (RefDet1F) [right=0.9cm of FBS, yshift=-0.575cm] {};
        \node (RefDet0T) [left=0.9cm of TH, yshift=0.575cm] {};
        \node (RefDet1T) [left=0.9cm of TH, yshift=-0.575cm] {};
        \node (RefTH0) [right=0.9cm of Det0, yshift=-0.5cm] {};
        \node (RefTH1) [right=0.9cm of Det1, yshift=0.5cm] {};
        \draw [-latex, thick] (SMF) -- (FBS);
        \draw [-latex, thick] (RefFBS0) -- (RefDet0F);
        \draw [-latex, thick] (RefFBS1) -- (RefDet1F);
        \draw [-latex, thick] (RefDet0T) -- (RefTH0);
        \draw [-latex, thick] (RefDet1T) -- (RefTH1);
        \end{tikzpicture}
        \caption{Second-order Temporal Correlation Function}
        \label{fig:g2config}
    \end{subfigure}

    \vspace{0.3cm}
    
    \begin{subfigure}[b]{.48\textwidth}
        \centering
        \begin{tikzpicture}[scale=0.35, font=\footnotesize]
        \node (SMF) [smallblock] {SMF};
        \node (Spect) [largeblock, right=0.9cm of SMF] {Ocean Optics Spectrometer};
        \draw [-latex, thick] (SMF) -- (Spect);
        \end{tikzpicture}
        \caption{Fluorescence Spectrum}
        \label{fig:spectrumconfig}
    \end{subfigure}
    \caption{Different configurations for the various types of analysis needed to identify and characterise regions of interest on the nanodiamond sample.  The abbreviations used here (not already introduced) are field-programmable gate array~(FPGA) and fibre beamsplitter~(FBS).  The final data acquisition device in each configuration is connected to a PC for data processing and recording.}
    \label{fig:configs}
\end{figure}

Before we proceed to demonstrate QRNG using photons emitted by NV centres in nanodiamonds, we first identify and characterise suitable regions of interest on our nanodiamond sample.  The configurations for these experiments are shown in \figref{fig:configs}.  To identify regions of interest, we perform a fluorescence scan.  To this end, the 2D GM (see \figref{fig:setup}) is used to simultaneously move both the pump beam and the capture region in discrete steps over a predefined scan area on the nanodiamond sample.  The SMF from the optical setup in \figref{fig:setup} is connected to a single-photon avalanche diode~(SPAD) detector (Excelitas SPCM-AQRH-15-FC), which is in turn connected to a FPGA board (see \figref{fig:scanconfig}).  This allows us to measure the photon detection rate at different points in the scan area as the 2D GM scans over the sample, which in turn enables us to assemble fluorescence scan images of our nanodiamond sample.

\begin{figure}
    \centering
    \begin{subfigure}[b]{.48\textwidth}
        {
        \subcaptionsetup{singlelinecheck=off}
        \hspace{-1.5cm}\subcaptionbox{Isolated Nanodiamonds\label{fig:nanodiamondscan}}
        {\hspace{1.5cm}\includegraphics[scale=0.54]{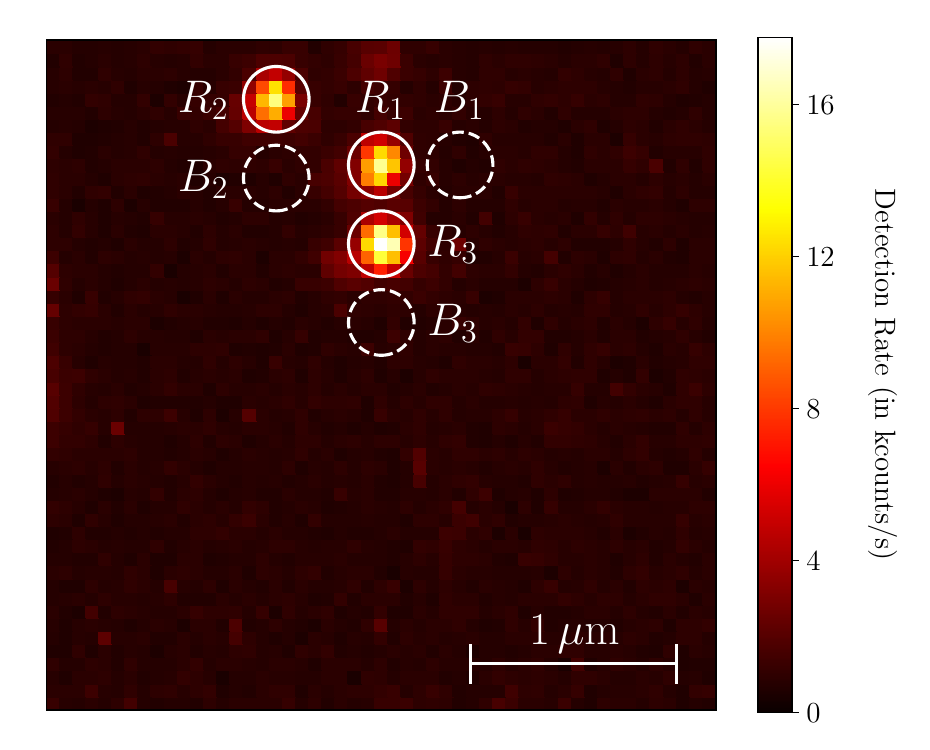}}
        }
    \end{subfigure}
    
    \vspace{0.25cm}
    
    \begin{subfigure}[b]{.48\textwidth}
        {
        \subcaptionsetup{singlelinecheck=off}
        \hspace{-1.5cm}\subcaptionbox{Clusters of Nanodiamonds\label{fig:clusterscan}}
        {\hspace{1.5cm}\includegraphics[scale=0.54]{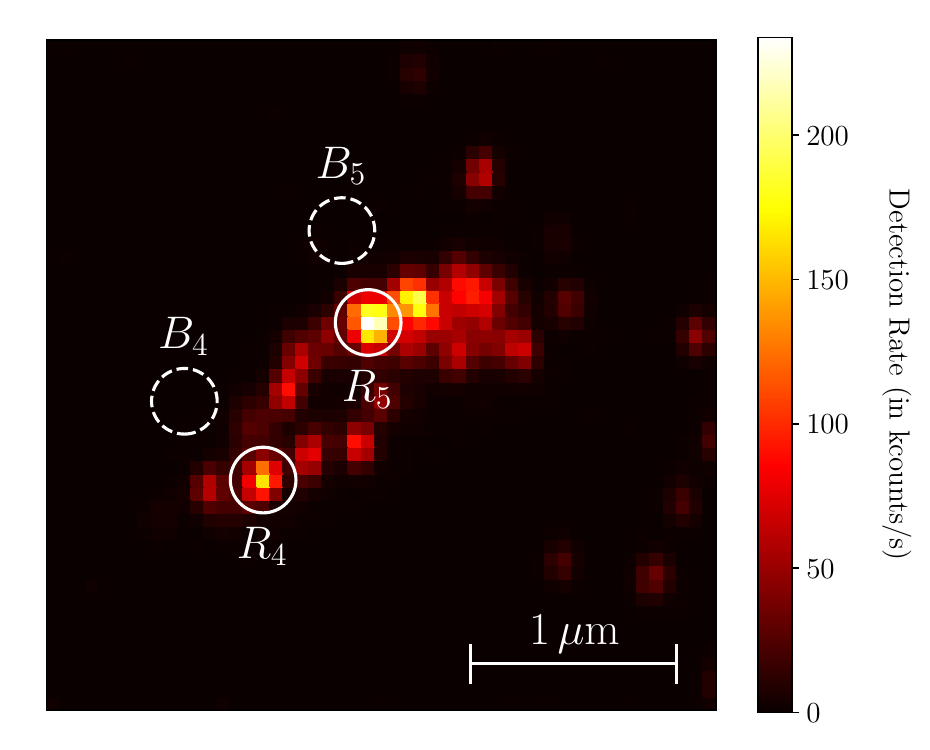}}
        }
    \end{subfigure}
    \caption{Fluorescence scan image of an area of the nanodiamond sample with (a)~Isolated Nanodiamonds and (b)~Clusters of Nanodiamonds.  The integration time for each pixel is $0.1\,\text{s}$.  The regions of interest are marked by solid white circles and are labelled $R_{1}$ to $R_{5}$.  The capture regions used to take background measurements for these respective regions are marked by dashed white circles and are labelled $B_{1}$ to $B_{5}$.}
    \label{fig:scan}
\end{figure}

\figref{fig:nanodiamondscan} shows a fluorescence scan image of an area with isolated nanodiamonds and \figref{fig:clusterscan} shows a scan of an area with clusters of nanodiamonds.  They are easily distinguishable, since the photon detection rate for isolated nanodiamonds is at least an order of magnitude lower than for clusters.  The five regions (three isolated nanodiamonds and two clusters of nanodiamonds) which we employ in our QRNG experiments (see \secref{sec:QRNGexperiment}) are marked by solid white circles on the scan images and are labelled $R_{1}$ to $R_{5}$.  In what follows, we first determine the number of NV centres present in each of these regions of interest.

To determine the number of NV centres present in a nanodiamond or cluster, we employ the second-order temporal correlation function, denoted by $g^{(2)}(\tau)$, where $\tau$ is a time delay.  It is related to the probability of a photon arriving at time $t$ and a second photon arriving at time $t + \tau$~\cite{g2interpretation}.  This approach therefore relies on temporal correlations in the emitted photons to infer the number of NV centres in a region.  For an ideal single-photon source, we would expect that $g^{(2)}(0)=0$~\cite{g2single}.

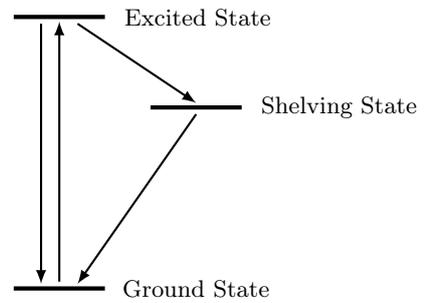
\begin{figure}
    \centering
    \begin{tikzpicture}[scale=1.2, font=\small]
    \draw[ultra thick] (0, 0) -- (1, 0);
    \draw[ultra thick] (0, 3) -- (1, 3);
    \draw[ultra thick] (1.5, 2) -- (2.5, 2);
    \draw[-latex, thick] (0.5, 0.075) -- (0.5, 2.95);
    \draw[-latex, thick] (0.3, 2.925) -- (0.3, 0.05);
    \draw[-latex, thick] (0.7, 2.925) -- (2, 2.05);
    \draw[-latex, thick] (2, 1.925) -- (0.7, 0.05);
    \node[] at (2, 0) {Ground State};
    \node[] at (2.02, 3) {Excited State};
    \node[] at (3.57, 2) {Shelving State};
    \end{tikzpicture}
    \caption{Three-level model developed for NV centres by Berthel \textit{et al.}~\cite{NVClevels}.  This model includes the ground state, the main excited state and a metastable shelving state.  The arrows indicate the allowed transitions taken into account in the model.}
    \label{fig:levels}
\end{figure}

Under the three-level model introduced by Kurtsiefer \textit{et al.}~\cite{g2threelevel} and developed further by Berthel \textit{et al.}~\cite{NVClevels} (see \figref{fig:levels}), the second-order temporal correlation function for a single, isolated NV centre is given by
\begin{equation}
g^{(2)}(\tau)=1-\beta e^{-\gamma_{1}|\tau|}+(\beta-1)e^{-\gamma_{2}|\tau|},
\label{eq:g2single}
\end{equation}
where $\gamma_{1}$ is the rate associated with transitions between the ground state and the main excited state (that is, the sum of the excitation rate and the spontaneous decay rate), $\gamma_{2}$ is the rate associated with decay from the main excited state to the ground state via a third metastable shelving state and $\beta$ is a parameter which quantifies the effect of shelving.  Since the shelving state is assumed to be metastable, it is assumed that $\gamma_{1} \gg \gamma_{2}$.  For $\text{NV}^0$ centres, the effect of shelving is generally negligible and $\beta \approx 1$.  For $\text{NV}^-$ centres, shelving has a more significant effect and $\beta$ tends to be larger, although at sufficiently low pump intensities, such as the pump intensity of $300\,\mu\text{W}$ used in our experiments, $\beta$ remains fairly close to $1$.  At higher pump intensities however, $\beta$ increases as the pump intensity increases.

It follows from \equatref{eq:g2single} that the second-order temporal correlation function for $N$ independent and identical NV centres is given by
\begin{equation}
g^{(2)}(\tau)=1-\frac{1}{N}\left(\beta e^{-\gamma_{1}|\tau|}-(\beta-1)e^{-\gamma_{2}|\tau|}\right).
\label{eq:g2multiple}
\end{equation}
This is an extension of a result which is well-established for simpler two-level models~\cite{g2twolevel1, g2twolevel2, g2twolevel3}, to the three-level model of Berthel \textit{et al.}~\cite{NVClevels}.  In a more realistic setting, where the NV centres are not all identical (possibly even a mix of $\text{NV}^0$ and $\text{NV}^-$ centres), the parameters $\gamma_{1}$, $\gamma_{2}$ and $\beta$ in \equatref{eq:g2multiple} can be interpreted as the average values of these parameters for the $N$ different NV centres, where $N$ represents the effective number taking into account the heterogeneous dipole orientation~\cite{g2twolevel1}.

In any experiment, some incoherent background light will be present in the collection and contaminate the ideal photon statistics.  To account for this, we follow the approach used in previous studies~\cite{NVClevels, NVCentanglement}, whereby \equatref{eq:g2multiple} becomes
\begin{equation}
g^{(2)}(\tau)=1-\frac{\rho^{2}}{N}\left(\beta e^{-\gamma_{1}|\tau|}-(\beta-1)e^{-\gamma_{2}|\tau|}\right),
\label{eq:g2background}
\end{equation}
where the parameter $\rho$ quantifies the effect of background contamination and is given by $\rho=\frac{S}{S+B}$.  Here, $S$ is the photon detection rate associated with the total signal for a region of interest and $B$ is the photon detection rate associated with the background light for that region.  To determine the value of $\rho$ for Region~$k$, we performed a separate experiment in which we obtained $S$ and $B$ by measuring the average photon detection rate in the capture regions labelled $R_{k}$ and $B_{k}$, respectively, on the fluorescence scan images in \figref{fig:scan}.

Accurately determining $\gamma_{2}$ through fitting requires data for $g^{(2)}(\tau)$ at long time delays $\tau$ (ideally at least up to $1\,\mu\text{s}$), which can be challenging and time-consuming to obtain experimentally.  It is also of little interest in the current context, since the primary objective is to determine the number of NV centres $N$.  We therefore employ results from previous experiments to approximate $\gamma_{2}$ in \equatref{eq:g2background}.  In experiments carried out with fluorescent nanodiamonds similar in size to the ones used here, it was found that $\gamma_{2} \approx \frac{\gamma_{1}}{20}$ for $\text{NV}^0$ centres~\cite{NVClevels}.  While this is not generally true for $\text{NV}^-$ centres, the result also holds approximately for $\text{NV}^-$ centres at pump intensities around $300\,\mu\text{W}$.  We therefore use this result to obtain
\begin{equation}
g^{(2)}(\tau)=1-\frac{\rho^{2}}{N}\left(\beta e^{-\gamma|\tau|}-(\beta-1)e^{-\gamma|\tau|/20}\right),
\label{eq:g2reparam}
\end{equation}
where the rate associated with transitions between the ground state and the main excited state is now simply denoted by $\gamma$.

\begin{figure*}
    \centering
    \begin{subfigure}[b]{.329\textwidth}
        \centering
        \includegraphics[scale=0.5]{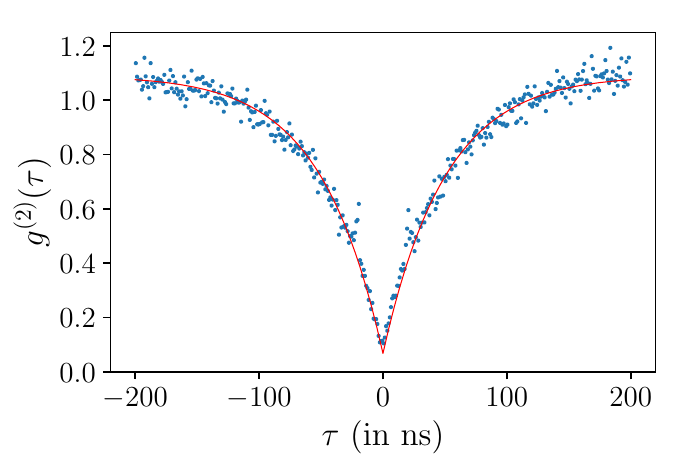}
        \caption{Region~1}
        \label{fig:g2curve1}
    \end{subfigure}
    \begin{subfigure}[b]{.329\textwidth}
        \centering
        \includegraphics[scale=0.5]{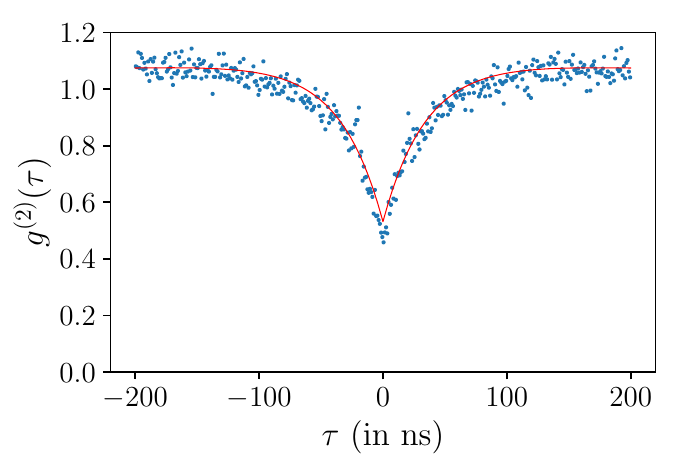}
        \caption{Region~2}
        \label{fig:g2curve2}
    \end{subfigure}
    \begin{subfigure}[b]{.329\textwidth}
        \centering
        \includegraphics[scale=0.5]{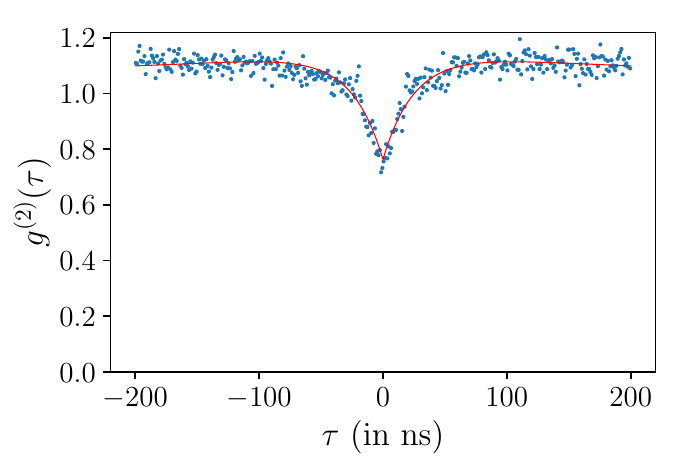}
        \caption{Region~3}
        \label{fig:g2curve3}
    \end{subfigure}

    \vspace{0.25cm}
    
    \begin{subfigure}[b]{.329\textwidth}
        \centering
        \includegraphics[scale=0.5]{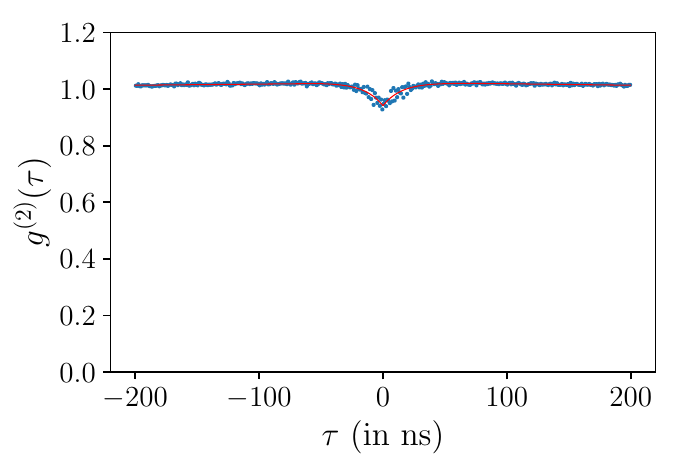}
        \caption{Region~4}
        \label{fig:g2curve4}
    \end{subfigure}
    \begin{subfigure}[b]{.329\textwidth}
        \centering
        \includegraphics[scale=0.5]{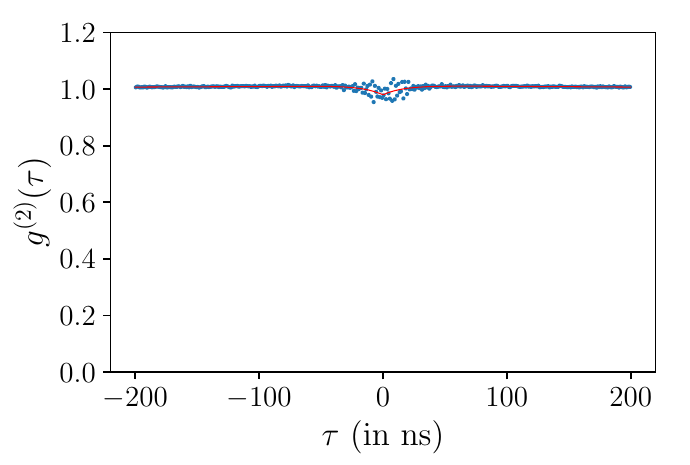}
        \caption{Region~5}
        \label{fig:g2curve5}
    \end{subfigure}
    \caption{The second-order temporal correlation functions for the different regions of interest.  The data points (shown in blue) are an average of 10 measurements, each with an integration time of $1200\,\text{s}$.  The best fit curves (see \equatref{eq:g2reparam}) are shown in red.}
    \label{fig:g2curves}
\end{figure*}

To measure $g^{(2)}(\tau)$ experimentally, we employ a multi-start multi-stop time-to-digital conversion method.  The configuration is shown in \figref{fig:g2config}.  The SMF from the optical setup in \figref{fig:setup} is connected to an input port of a $50$/$50$ fibre beamsplitter~(FBS) (Thorlabs TW670R5F2).  Each of the two output ports of the FBS is connected to a SPAD detector which is similar to the one that was used for the fluorescence scan.  Each of these detectors is in turn connected to a channel of a time-to-digital converter (PicoQuant TimeHarp 260 PICO), which measures the arrival times of photons at each detector to a precision of $25\,\text{ps}$.  The data is processed in real time on a PC using the QuCoa software.  In particular, values of the (normalised) second-order temporal correlation function for different time delays $\tau$ are obtained using
\begin{equation}
g^{(2)}(\tau)=\frac{c(\tau)}{C_{0}C_{1} \Delta\tau T_{\text{tot}}},
\label{eq:g2exp}
\end{equation}
where $T_{\text{tot}}=1200\,\text{s}$ is the total integration time for a measurement, $\Delta\tau=1\,\text{ns}$ is the width of the coincidence time window, $C_{k}$ is the total number of photon counts at Detector~$k$ during the integration time $T_{\text{tot}}$ and $c(\tau)$ is the total number of coincident counts, within the coincidence time window $\Delta\tau$, with a time delay $\tau$ between Detector~$0$ and Detector~$1$, during the integration time $T_{\text{tot}}$.  We fitted \equatref{eq:g2reparam} to the experimentally measured values from \equatref{eq:g2exp}, treating $N$ as an integer fitting parameter, and $\gamma$ and $\beta$ as real fitting parameters, while a separate experiment was performed to determine $\rho$.

The values of $g^{(2)}(\tau)$ for each region of interest are plotted in \figref{fig:g2curves}.  In these plots, the experimental data points are shown in blue and the fitted curves are shown in red.  The fitting parameters obtained for the different regions of interest are given in \tabref{tab:g2params}, along with the experimentally measured values of $\rho$.  Through fitting, we were able to infer that the nanodiamond in Region~1 contains a single NV centre, while the nanodiamonds in Regions~2 and~3 contain two and four NV centres, respectively.  Furthermore, we inferred that the cluster of nanodiamonds in Region~4 contains around 17 NV centres and that the cluster in Region~5 contains just under 50 NV centres.  To investigate the presence of $\text{NV}^0$ and $\text{NV}^-$ centres in these regions, we measured their fluorescence spectra (see \appendref{append:spectra}).  The results suggest that a single $\text{NV}^0$ centre is present in Region~1 and that two $\text{NV}^0$ centres are present in Region~2, while both $\text{NV}^0$ and $\text{NV}^-$ centres are present in the other regions.  This is consistent with the results in \tabref{tab:g2params}.  In particular, we see that $\beta \approx 1$ for Regions~1 and~2, as expected for $\text{NV}^0$ centres, while the average value of $\beta$ is slightly larger for Regions~3, 4 and~5, due to the presence of $\text{NV}^-$ centres.

\begin{table}[]
\begin{tabular}{|l|l|l|l|l|l|}
\hline
\textbf{Region} & $\boldsymbol{N}$ & $\boldsymbol{\gamma\ \text{\textbf{(in ns}}^{-1}}\text{\textbf{)}}$ & $\boldsymbol{\beta}$ & $\boldsymbol{\rho}$ \\ \hline
1          & 1          & 0.020$\pm$0.001             & 1.126$\pm$0.022             & 0.96531$\pm$0.00245             \\ \hline
2          & 2          & 0.029$\pm$0.012             & 1.218$\pm$0.047             & 0.96816$\pm$0.00185             \\ \hline
3          & 4          & 0.040$\pm$0.008             & 1.623$\pm$0.055             & 0.97251$\pm$0.00229             \\ \hline
4          & 17         & 0.066$\pm$0.003             & 1.455$\pm$0.050             & 0.99708$\pm$0.00008             \\ \hline
5          & 49         & 0.063$\pm$0.004             & 1.671$\pm$0.070             & 0.99839$\pm$0.00005             \\ \hline
\end{tabular}
\caption{The second-order temporal correlation function parameters (see \equatref{eq:g2reparam}) obtained for the different regions of interest.  We treated $N$ as an integer fitting parameter and $\gamma$ and $\beta$ as real fitting parameters, and performed a separate experiment to measure $\rho$.  The estimated values of these parameters are an average of 10 repetitions and the uncertainties are given by the standard deviation.}
\label{tab:g2params}
\end{table}

\section{QRNG}\label{sec:QRNG} 

\subsection{The time-of-arrival scheme}\label{sec:QRNGscheme}

\begin{figure}
    \centering
    \begin{subfigure}[b]{.48\textwidth}
        \centering
        \begin{tikzpicture}[scale=0.8, font=\scriptsize]
        \draw[very thick] (0, 0) -- (6, 0);
        \draw[very thick] (0, -0.3) -- (0, 0.3);
        \draw[very thick] (1, -0.3) -- (1, 0.3);
        \draw[very thick] (2, -0.3) -- (2, 0.3);
        \draw[very thick] (3, -0.3) -- (3, 0.3);
        \draw[very thick] (4, -0.3) -- (4, 0.3);
        \draw[very thick] (5, -0.3) -- (5, 0.3);
        \draw[very thick] (6, -0.3) -- (6, 0.3);
        \draw (0, 0.8) -- (0.6, 0.8) -- (0.6, 1.2) -- (0.8, 1.2) -- (0.8, 0.8) -- (3.4, 0.8) -- (3.4, 1.2) -- (3.6, 1.2) -- (3.6, 0.8) -- (4.7, 0.8) -- (4.7, 1.2) -- (4.9, 1.2) -- (4.9, 0.8) -- (6, 0.8);
        \draw[dashed] (0.6, 0) -- (0.6, 0.8);
        \draw[dashed] (3.4, 0) -- (3.4, 0.8);
        \draw[dashed] (4.7, 0) -- (4.7, 0.8);
        \draw[latex-latex] (0, 0.3) -- (0.6, 0.3);
        \draw[latex-latex] (3, 0.3) -- (3.4, 0.3);
        \draw[latex-latex] (4, 0.3) -- (4.7, 0.3);
        \node[] at (0.5, -0.18) {$T$};
        \node[] at (1.5, -0.18) {$T$};
        \node[] at (2.5, -0.18) {$T$};
        \node[] at (3.5, -0.18) {$T$};
        \node[] at (4.5, -0.18) {$T$};
        \node[] at (5.5, -0.18) {$T$};
        \node[] at (0.4, 1.1) {$t_1$};
        \node[] at (3.2, 1.1) {$t_2$};
        \node[] at (4.5, 1.1) {$t_3$};
        \node[] at (6.9, 0) {Reference};
        \node[] at (7.2, 0.85) {Arrival Times};
        \end{tikzpicture}
        \caption{Time-of-arrival Scheme}
        \label{fig:toascheme}
    \end{subfigure}

    \vspace{0.35cm}
    
    \begin{subfigure}[b]{.48\textwidth}
        \centering
        \begin{tikzpicture}[scale=0.35, font=\footnotesize]
        \node (SMF) [smallblock] {SMF};
        \node (Det) [mediumblock, right=0.9cm of SMF] {Detector};
        \node (TH) [medlargeblock, right=0.9cm of Det] {PicoQuant TimeHarp};
        \draw [-latex, thick] (SMF) -- (Det);
        \draw [-latex, thick] (Det) -- (TH);
        \end{tikzpicture}
        \caption{Experimental Configuration}
        \label{fig:toaconfig}
    \end{subfigure}
    \caption{The time-of-arrival scheme for QRNG.  (a)~Variation of the time-of-arrival scheme first proposed by Nie \textit{et al.}~\cite{QRNGtoa5}, in which random numbers are determined by the arrival times of photons at a detector relative to an external time reference.  (b)~The configuration for our experimental implementation of the time-of-arrival scheme.}
    \label{fig:toa}
\end{figure}
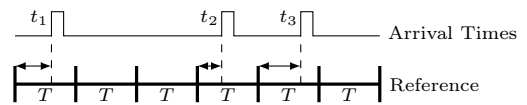
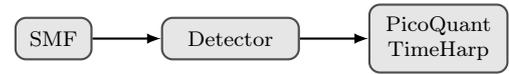

The spontaneous emission of a photon from a NV centre when it spontaneously decays from its main excited state to its ground state is an inherently random quantum mechanical process.  We generate random numbers from the arrival times of photons emitted by NV centres in the regions of interest we identified and characterised in \secref{sec:characterisation}.  To this end, we implement the variation of the time-of-arrival scheme for QRNG first proposed by Nie \textit{et al.}~\cite{QRNGtoa5} (see \figref{fig:toascheme}).  In this variation, random numbers are determined by the arrival times of photons at a detector relative to an external time reference, which is divided into periodic time intervals of length $T$.  Each of these periodic time intervals is broken up into $M$ equal subintervals or bins, each of which have a width of $\tau=\frac{T}{M}$.  For a coherent source, it has been shown that a photon arriving during one of these periodic time intervals of length $T$ arrives in any one of the $M$ bins with a probability of $\frac{1}{M}$, under the assumption that at most one photon can arrive during a time interval of length $T$~\cite{QRNGtoa5, QRNGtoa8}.  In \appendref{append:uniformity}, we show that this result holds for any light source, so long as at most one photon arrival event can occur during a time interval of length $T$.  In this ideal case, the indices of the bins in which photons arrive are uniformly distributed integers in the range $[1,\,M]$.

The configuration for our experimental implementation of the time-of-arrival scheme with photons emitted by NV centres is shown in \figref{fig:toaconfig}.  The SMF from the optical setup in \figref{fig:setup} is connected to a SPAD detector which is similar to the ones that were used in the experiments in \secref{sec:characterisation}.  These detectors have a dead time of $24\,\text{ns}$ and a maximum dark count rate of $50\,\text{counts/s}$.  The SPAD detector is connected to a channel of the same PicoQuant TimeHarp time-to-digital converter that was used for the second-order correlation measurements in \secref{sec:characterisation}.  The arrival times of photons at the detector are recorded on a PC, to a precision of $25\,\text{ps}$, and an offline Python script is used to convert the recorded photon arrival times into binary digits (bits) which correspond to the indices of the bins in which the photon arrivals occurred.  In our implementation, we set $T=12.8\,\text{ns}$ and $M=2^{8}=256$, and so $\tau=50\,\text{ps}$.  This allows us to obtain an $8$-bit string (or a byte) from each recorded photon arrival time.  Furthermore, since $T$ is less than the dead time of the detector, at most one photon arrival can be registered during each time interval of length $T$, and since $\tau$ is greater than the timing resolution of the TimeHarp, the individual bins can be properly resolved.

In any experimental implementation of the time-of-arrival scheme, device imperfections will be present, which could introduce bias or correlations and degrade the quality of the random numbers generated by the implementation.  These include detector imperfections and source imperfections.  Detector imperfections which have been considered in previous implementations~\cite{QRNGtoa5, QRNGtoa8} include the detector dead time, which reduces the efficiency of the detector, and detector dark counts.  In our experiments, the photon detection rate ranges from about $20\,\text{kcounts/s}$ for the single NV centre in Region~1 to about $600\,\text{kcounts/s}$ for the cluster of NV centres in Region~5.  Hence, even at the highest photon detection rate, the average time interval between consecutive photon detections is around $2\,\mu\text{s}$, which is two orders of magnitude greater than the detector dead time.  As such, almost all the photons arriving at the detector will be detected, and so the detector efficiency (in terms of its ability to resolve two separate detection events) is very close to unity in our experiments.  Furthermore, even the lowest photon detection rate is three orders of magnitude greater than the maximum dark count rate, and so the effect of dark counts will also be negligible.  The timing jitter of $350\,\text{ps}$ of the SPAD detector introduces a small amount of additional random noise corresponding to $\pm 3$ time bins for the arrival time.  However, previous experiments have found that this does not affect the uniformity or quality of the random numbers generated~\cite{QRNGtoa5, QRNGtoa8}.  The effect of source imperfections, in the form of multi-photon emission, has been extensively analysed for a coherent source~\cite{QRNGtoa5, QRNGtoa8}.  In the next section, we present a detailed theoretical analysis of the effect of source imperfections for a single-photon emitter, as well as for multiple independent single-photon emitters, and apply the general theory to our experimental implementation of the time-of-arrival scheme with photons emitted by NV centres.

\subsection{Theoretical background and analysis}\label{sec:QRNGtheory}

\subsubsection{A single-photon emitter}\label{sec:QRNGtheorysingle}

For any light source, we show in \appendref{append:uniformity} that uniform random numbers will be generated by an implementation of the time-of-arrival scheme, provided that no multi-photon events can occur during a time interval of length $T$.  However, even for an ideal single-photon source (which never emits multiple photons simultaneously) there is a non-zero probability that when driven by a CW pump laser multiple photons will be emitted during any finite time interval due to the intrinsic lifetime of the source and so no light source can truly satisfy this assumption.  Here we analyse the effect of multi-photon events on the uniformity and quality of the random numbers generated by an implementation of the time-of-arrival scheme for a three-level single-photon emitter (see \figref{fig:levels}) with background contamination.  Let $P_{t}^{(1)}(n)$ denote the probability of $n$ photons from a single-photon emitter arriving during a time interval of length $t$.  We follow the approach of Ref.~\cite{NVCentanglement} and use a bound from Ref.~\cite{probabilities} to show that the photon number probabilities for a single-photon emitter can be approximated by (see \appendref{append:probabilities})
\begin{align*}
P_{t}^{(1)}(0) &= 1 - \mu_{t} + \frac{1}{2}\mu_{t}^{2}g_{t}^{(2)}(0), \\
P_{t}^{(1)}(1) &= \mu_{t} - \mu_{t}^{2}g_{t}^{(2)}(0), \\
P_{t}^{(1)}(2) &= \frac{1}{2}\mu_{t}^{2}g_{t}^{(2)}(0),
\end{align*}
and $P_{t}^{(1)}(n) = 0$ for $n \ge 3$, under the assumption that $t \ll \frac{1}{\gamma_{1}}$.  In these expressions, $\mu_{t}=\lambda t$ is the mean number of photons arriving during a time interval of length $t$, with $\lambda$ the time-independent mean photon flux, and $g_{t}^{(2)}(0)$ is the detected value of the second-order correlation function measured over a finite time interval of length $t$ around zero, which is given by
\begin{align}
g_{t}^{(2)}(0) &= 1-\frac{2\rho^{2}}{\gamma_{1}^{2}\gamma_{2}^{2}t^{2}}\left(\beta\gamma_{2}^{2}\left(e^{-\gamma_{1}t}+\gamma_{1}t-1\right)\right. \nonumber \\
&\quad \left.-(\beta-1)\gamma_{1}^{2}\left(e^{-\gamma_{2}t}+\gamma_{2}t-1\right)\right),
\label{eq:g2detected}
\end{align}
for a three-level single-photon emitter with background contamination~\cite{NVCentanglement}.

To analyse the effect of multi-photon events on the uniformity and quality of the random numbers generated by an implementation of the time-of-arrival scheme, we consider a time interval of length $T$ divided into $M$ bins of equal width, $\tau=\frac{T}{M}$.  Since a random number is generated if and only if a photon arrives during time $T$, we assume that at least one photon arrives during time $T$, but place no further restrictions on the total number of photons arriving during time $T$.  Note that when multiple photons can arrive during time $T$, the generated random number is the index of the first bin in which a photon arrives.  For a single-photon emitter, the probability that the first photon arriving during time $T$ arrives in an arbitrary bin $i$ is given by
\begin{equation}
p_{i} = \frac{(P_{\tau}^{(1)}(0))^{i-1}(P_{\tau}^{(1)}(1)+P_{\tau}^{(1)}(2))}{1-(P_{\tau}^{(1)}(0))^{M}}.
\label{eq:pisingle}
\end{equation}
This expression can be understood as follows.  The numerator is the probability that no photons arrive in the first $i-1$ bins and a non-zero number of photons (either one or two, since $P_{\tau}^{(1)}(n) = 0$ for $n \ge 3$) arrive in bin~$i$, which is precisely the probability that the first photon arriving during time $T$ arrives in bin~$i$.  Since we do not restrict the total number of photons arriving during time $T$, the $M-i$ bins after bin~$i$ do not enter into the expression, as any number of photons can arrive in any of these bins with no effect on bin~$i$ being the first bin in which a photon arrives.  The denominator is the normalisation constant for the probabilities $p_{i}$ and is needed since we assume that at least one photon arrives during time $T$, and so the case where all $M$ bins are empty needs to be eliminated from the sample space.  The denominator is therefore one minus the probability that no photons arrive in any of the $M$ bins.

Just as for a coherent source~\cite{QRNGtoa5, QRNGtoa8}, the quality of the random numbers generated with a single-photon emitter can be quantified by the min-entropy per bit, which is defined by
\begin{equation}
H_{\infty} = -\log_{M}(\max_{i}p_{i}).
\label{eq:minentdef}
\end{equation}
To find the maximum of the probabilities $p_{i}$, we note that $(P_{\tau}^{(1)}(0))^{i} \ge (P_{\tau}^{(1)}(0))^{i+1}$, since $0 \le P_{\tau}^{(1)}(0) \le 1$, and so $p_{i} \ge p_{i+1}$ for all $i$.  Hence, $p_{1} \ge p_{i}$ for all $i$, which means that the first bin $i=1$ has the highest probability of being the first bin in which a photon arrives during the time interval of length $T$.  It therefore follows from Eqs.~(\ref{eq:pisingle}) and~(\ref{eq:minentdef}) that
\begin{equation}
H_{\infty} = -\log_{M}\left(\frac{P_{\tau}^{(1)}(1)+P_{\tau}^{(1)}(2)}{1-(P_{\tau}^{(1)}(0))^{M}}\right),
\label{eq:minentsingle}
\end{equation}
for a single-photon emitter.  In contrast, the min-entropy per bit is given by
\begin{equation}
H_{\infty} = 1 + \log_{M}(1-e^{-\lambda T}) - \log_{M}(\lambda T),
\label{eq:minentcoherent}
\end{equation}
for a coherent source~\cite{QRNGtoa5, QRNGtoa8}.  Hence, for a coherent source, the min-entropy depends only on $T$ and the mean photon flux $\lambda$, while the min-entropy for a single-photon emitter also depends on the second-order temporal correlation function parameters ($\gamma_{1}$, $\gamma_{2}$, $\beta$ and $\rho$).

\begin{figure}
    \centering
    \begin{subfigure}[b]{.48\textwidth}
        {
        \subcaptionsetup{singlelinecheck=off}
        \hspace{1.2cm}\subcaptionbox{$p_{i}$ versus $i$\label{fig:probabilitysingle}}
        {\hspace{-1.2cm}\includegraphics[scale=0.5]{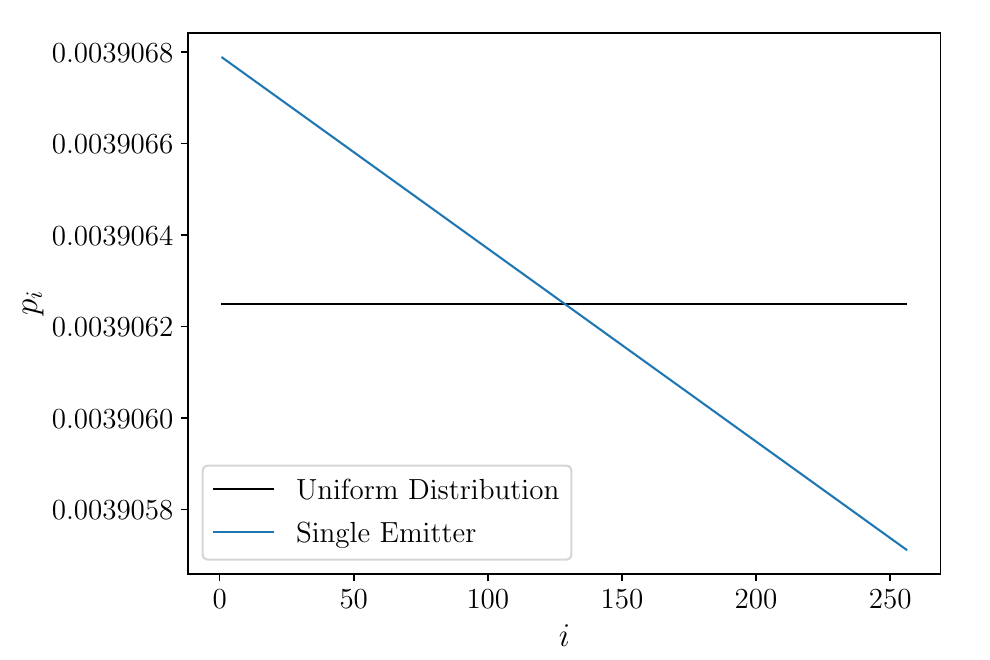}}
        }
    \end{subfigure}
    
    \vspace{0.25cm}
    
    \begin{subfigure}[b]{.48\textwidth}
        {
        \subcaptionsetup{singlelinecheck=off}
        \hspace{0.8cm}\subcaptionbox{$H_{\infty}$ versus $\lambda$\label{fig:entropysingle}}
        {\hspace{-0.8cm}\includegraphics[scale=0.5]{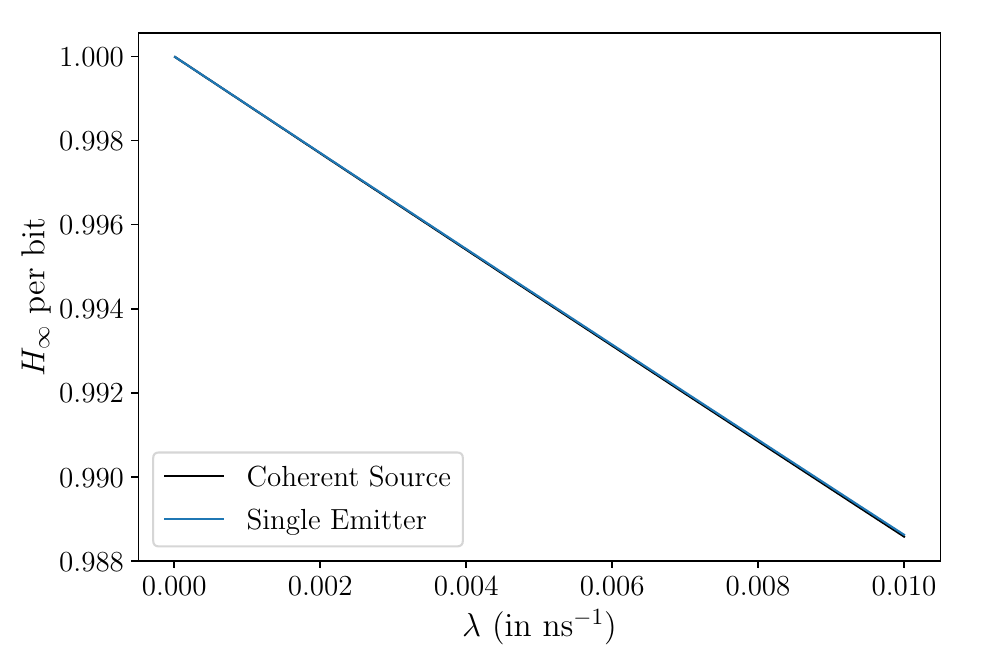}}
        }
    \end{subfigure}
    \caption{Theoretical analysis of QRNG based on the arrival times of photons emitted by the single NV centre in Region~1.  The second-order temporal correlation function parameters ($\gamma_{1}=\gamma$, $\gamma_{2}=\frac{\gamma}{20}$, $\beta$ and $\rho$) used for this analysis are given in \tabref{tab:g2params}.  (a)~$p_{i}$ versus $i$ for both the ideal uniform distribution and the single NV centre in Region~1, with $\lambda=0.0000216\,\text{ns}^{-1}$.  (b)~$H_{\infty}$ versus $\lambda$ for both a coherent source and the single NV centre in Region~1.}
    \label{fig:single}
\end{figure}

We now apply our general theory for a three-level single-photon emitter with background contamination to the single NV centre in Region~1.  A plot of $p_{i}$ versus $i$ is shown in \figref{fig:probabilitysingle}.  We see that the probability distribution calculated for the single NV centre in Region~1 deviates only marginally from the ideal uniform distribution.  In \figref{fig:entropysingle}, we plot $H_{\infty}$ versus $\lambda$ for the single NV centre in Region~1 as well as for a coherent source.  For a given $\lambda$, the min-entropy obtained for the single NV centre in Region~1 is approximately the same as for a coherent source.  Note that we do not plot beyond $\lambda=0.01\,\text{ns}^{-1}$, since for a single-photon emitter $\lambda$ is fundamentally limited by $\gamma_{1}$, the rate associated with transitions between the ground state and the main excited state of the emitter, which is typically of the order $0.01\,\text{ns}^{-1}$ (see \tabref{tab:g2params}).  This is in contrast with a coherent source, where $\lambda$ can be increased arbitrarily by increasing the intensity of the laser.  We also numerically investigated the effect of variations in the second-order temporal correlation function parameters ($\gamma_{1}$, $\gamma_{2}$, $\beta$ and $\rho$) on the probability distribution and the min-entropy for a single-photon emitter.  We found that variations in these parameters have a negligible effect on both the probability distribution and the min-entropy.

\subsubsection{Multiple single-photon emitters}\label{sec:QRNGtheorymultiple}

Our theory for a single-photon emitter can easily be generalised to the case where multiple single-photon emitters are present.  We simply need to find the probability of $n$ photons from $N$ single-photon emitters arriving during a time interval of length $t$, which we denote by $P_{t}^{(N)}(n)$.  To this end, we assume that the $N$ single-photon emitters are statistically independent and identical, and use the photon number probabilities for a single-photon emitter.  Since $P_{t}^{(1)}(n) = 0$ for $n \ge 3$, that is each of the $N$ single-photon emitters can emit at most two photons during a time interval of length $t$, it follows that $P_{t}^{(N)}(n) = 0$ for $n > 2N$.  For $0 \le n \le 2N$, we have
\begin{align}
P_{t}^{(N)}(n) &= \sum_{k=\max{(0,\,n-N)}}^{\left\lfloor\frac{n}{2}\right\rfloor}\frac{N!}{(N-n+k)!\,(n-2k)!\,k!} \nonumber \\
&\quad \times\left(P_{t}^{(1)}(0)\right)^{N-n+k}\left(P_{t}^{(1)}(1)\right)^{n-2k} \nonumber \\
&\quad \times\left(P_{t}^{(1)}(2)\right)^{k},
\label{eq:Pn}
\end{align}
where $0! = 1$ and $p^{0} = 1$, for $p > 0$, by convention.

In \equatref{eq:Pn}, we sum over the possible number of emitters, $k$, undergoing two-photon emission during the time interval of length $t$.  Note that if $k$ emitters emit two photons (third line), then $n-2k$ emitters must emit one photon and $N-n+k$ emitters must emit no photons (second line) in order for a total of $n$ photons to be emitted by the $N$ emitters during time $t$.  The bounds of the summation can be understood as follows.  Clearly we cannot have more than $\left\lfloor\frac{n}{2}\right\rfloor$ emitters emitting two photons, as this would result in the total number of photons emitted during time $t$ exceeding $n$.  Hence, the upper bound of the summation is $k = \left\lfloor\frac{n}{2}\right\rfloor$.  For a given $n$ and $N$ considered for $P_{t}^{(N)}(n)$, if there are fewer photons than emitters, that is when $0 \le n \le N$, it is not necessary for any of the emitters to emit two photons during time $t$, and so in this case the summation starts at $k = 0$.  On the other hand, if there are more photons than emitters, that is when $N < n \le 2N$, some of the emitters have to undergo two-photon emission during time $t$, as it is not possible for the number of photons to exceed the number of emitters otherwise.  In this case the summation starts at $k = n-N$, since the minimum number of emitters which need to emit two photons is equal to the number of photons exceeding the number of emitters.  It therefore follows that the lower bound of the summation is $k = \max{(0,\,n-N)}$.  The prefactor accounts for the distinct permutations of the emitters, which are identical and therefore totally indistinguishable except by the number of photons they emit.  In particular, there are $N$ emitters in total and the $N-n+k$ emitters emitting no photons are indistinguishable, the $n-2k$ emitters emitting one photon are also indistinguishable and the $k$ emitters emitting two photons are indistinguishable as well.

\begin{figure}
    \centering
    \begin{subfigure}[b]{.48\textwidth}
        {
        \subcaptionsetup{singlelinecheck=off}
        \hspace{1.15cm}\subcaptionbox{$p_{i}$ versus $i$\label{fig:probabilitymultiple}}
        {\hspace{-1.15cm}\includegraphics[scale=0.5]{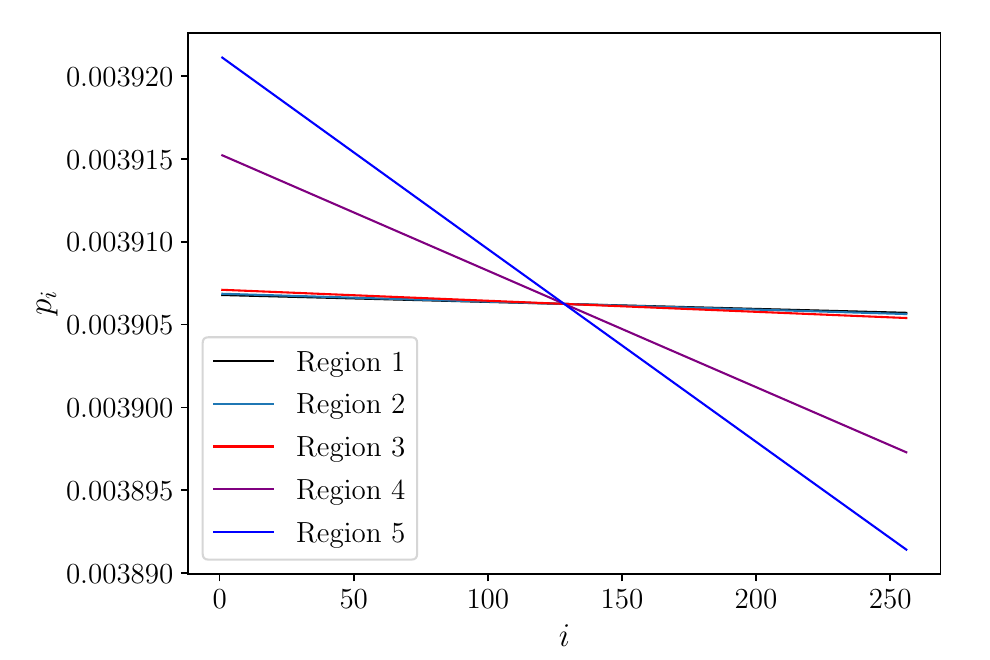}}
        }
    \end{subfigure}
    
    \vspace{0.25cm}
    
    \begin{subfigure}[b]{.48\textwidth}
        {
        \subcaptionsetup{singlelinecheck=off}
        \hspace{0.75cm}\subcaptionbox{$H_{\infty}$ versus $\lambda$\label{fig:entropymultiple}}
        {\hspace{-0.75cm}\includegraphics[scale=0.5]{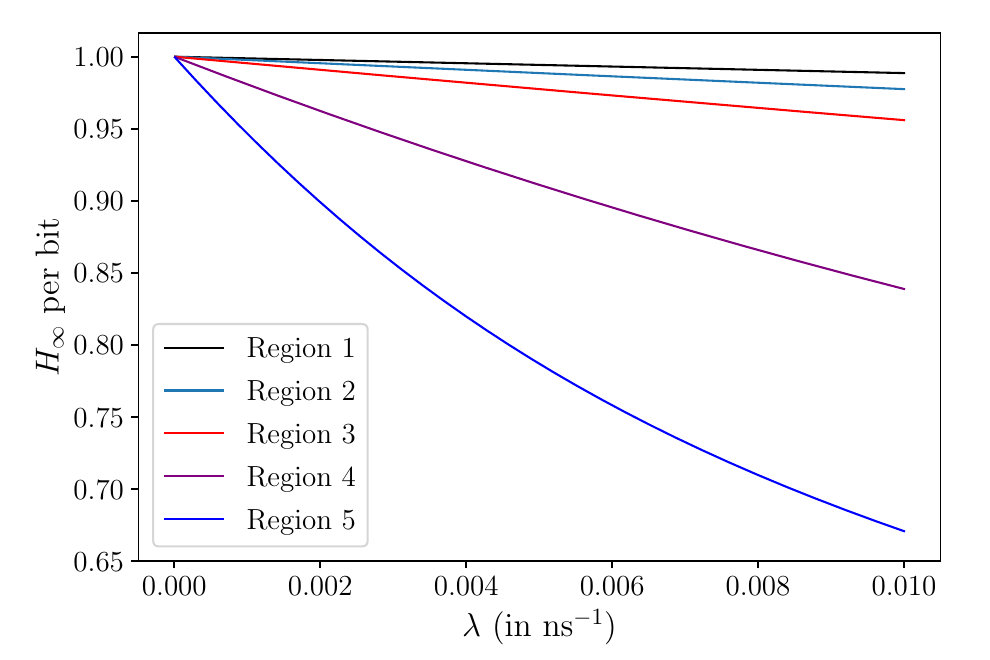}}
        }
    \end{subfigure}
    \caption{Theoretical analysis of QRNG based on the arrival times of photons emitted by NV centres in the different regions of interest.  The second-order temporal correlation function parameters ($\gamma_{1}=\gamma$, $\gamma_{2}=\frac{\gamma}{20}$, $\beta$ and $\rho$) used for this analysis are given in \tabref{tab:g2params}.  (a)~$p_{i}$ versus $i$ for the different regions of interest, with the values of $\lambda$ in \tabref{tab:minent} used.  (b)~$H_{\infty}$ versus $\lambda$ for the different regions of interest.}
    \label{fig:multiple}
\end{figure}

For multiple single-photon emitters, the probability that the first photon arriving during a time interval of length $T$, divided into $M$ bins of equal width, $\tau=\frac{T}{M}$, arrives in an arbitrary bin $i$ is given by
\begin{equation}
p_{i} = \frac{(P_{\tau}^{(N)}(0))^{i-1}}{1-(P_{\tau}^{(N)}(0))^{M}}\sum_{\ell=1}^{2N}P_{\tau}^{(N)}(\ell).
\label{eq:pimultiple}
\end{equation}
This can be understood by the explanation given in \secref{sec:QRNGtheorysingle} for a single-photon emitter.  Furthermore, the first bin $i=1$ once again has the highest probability of being the first bin in which a photon arrives.  Substituting into \equatref{eq:minentdef}, we obtain
\begin{equation}
H_{\infty} = -\log_{M}\left(\frac{1}{1-(P_{\tau}^{(N)}(0))^{M}}\sum_{\ell=1}^{2N}P_{\tau}^{(N)}(\ell)\right),
\label{eq:minentmultiple}
\end{equation}
for $N$ statistically independent and identical single-photon emitters.  Just as for a single emitter, the min-entropy depends on $\lambda$, which now denotes the mean photon flux per emitter, not the total mean photon flux for the $N$ emitters, as well as on the second-order temporal correlation function parameters ($\gamma_{1}$, $\gamma_{2}$, $\beta$ and $\rho$).

\begin{table}[]
\begin{tabular}{|l|l|l|l|}
\hline
\textbf{Region} & $\boldsymbol{\lambda\ \text{\textbf{(in ns}}^{-1}}\text{\textbf{)}}$ & $\boldsymbol{H_{\infty}\ \text{\textbf{per bit}}}$ \\ \hline
1          & 0.0000216          & 0.999975             \\ \hline
2          & 0.0000123          & 0.999972             \\ \hline
3          & 0.0000086          & 0.999961             \\ \hline
4          & 0.0000212          & 0.999586             \\ \hline
5          & 0.0000122          & 0.999314             \\ \hline
\end{tabular}
\caption{The calculated values of the min-entropy, $H_{\infty}$, and the experimentally measured values of the mean photon flux per emitter, $\lambda$, for the different regions of interest.  The second-order temporal correlation function parameters ($\gamma_{1}=\gamma$, $\gamma_{2}=\frac{\gamma}{20}$, $\beta$ and $\rho$) used in the min-entropy calculation are given in \tabref{tab:g2params}.}
\label{tab:minent}
\end{table}

\begin{table}[]
\begin{tabular}{|l|l|l|}
\hline
\textbf{Region} & \textbf{Speed (in Mbits/s)} \\ \hline
1          & 0.173          \\ \hline
2          & 0.197          \\ \hline
3          & 0.274          \\ \hline
4          & 2.88           \\ \hline
5          & 4.77           \\ \hline
\end{tabular}
\caption{The random number generation rates (speeds) achieved for the different regions of interest.}
\label{tab:rates}
\end{table}

\begin{table}[]
\begin{tabular}{|l|l|l|l|}
\hline
\textbf{Region} & \textbf{Frequency of 0} & \textbf{Frequency of 1} \\ \hline
1          & 0.500012          & 0.499988             \\ \hline
2          & 0.500000          & 0.500000             \\ \hline
3          & 0.500002          & 0.499998             \\ \hline
4          & 0.500000          & 0.500000             \\ \hline
5          & 0.500003          & 0.499997             \\ \hline
\end{tabular}
\caption{The relative frequency of 0 and 1 in the different $800\,\text{Mbit}$ samples generated from the arrival times of photons emitted by NV centres in the different regions of interest.}
\label{tab:relfreq}
\end{table}

\begin{figure*}
    \centering
    \begin{subfigure}[b]{.329\textwidth}
        \centering
        \includegraphics[scale=0.5]{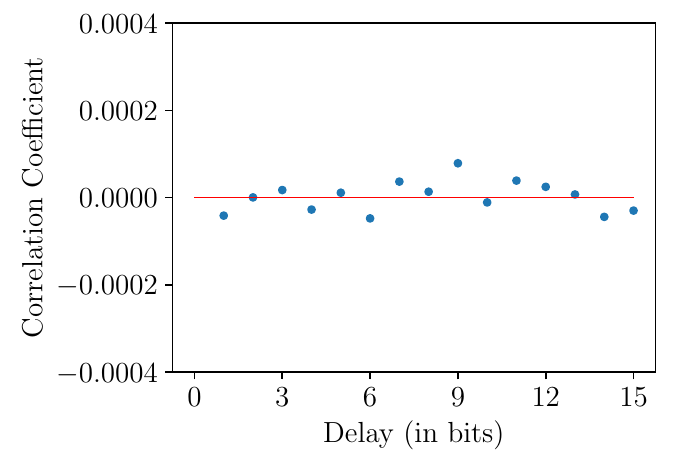}
        \caption{Region~1}
        \label{fig:pcoff1}
    \end{subfigure}
    \begin{subfigure}[b]{.329\textwidth}
        \centering
        \includegraphics[scale=0.5]{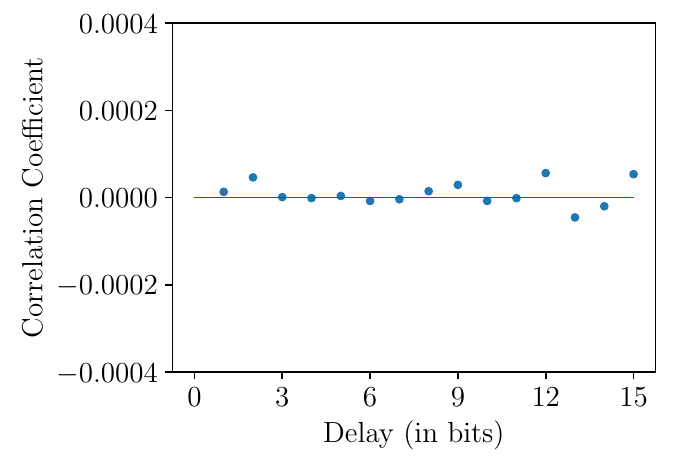}
        \caption{Region~2}
        \label{fig:pcoff2}
    \end{subfigure}
    \begin{subfigure}[b]{.329\textwidth}
        \centering
        \includegraphics[scale=0.5]{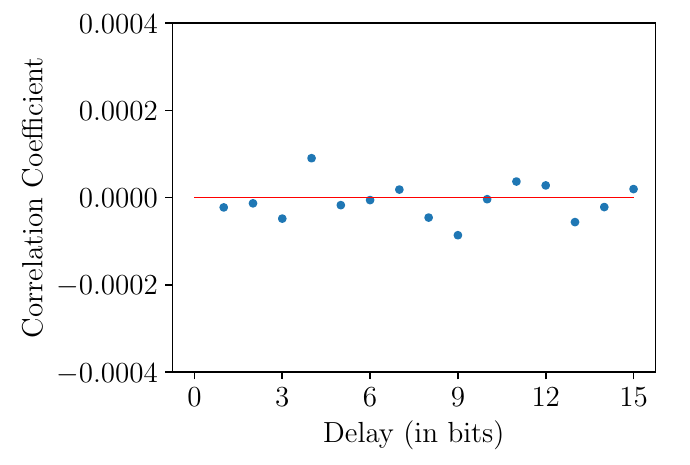}
        \caption{Region~3}
        \label{fig:pcoff3}
    \end{subfigure}

    \vspace{0.3cm}
    
    \begin{subfigure}[b]{.329\textwidth}
        \centering
        \includegraphics[scale=0.5]{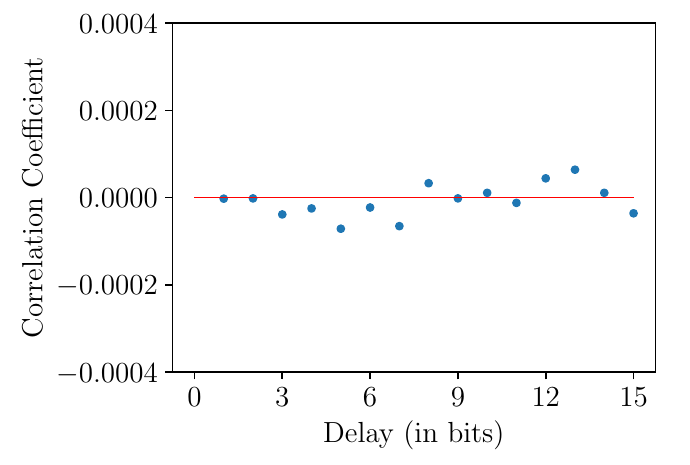}
        \caption{Region~4}
        \label{fig:pcoff4}
    \end{subfigure}
    \begin{subfigure}[b]{.329\textwidth}
        \centering
        \includegraphics[scale=0.5]{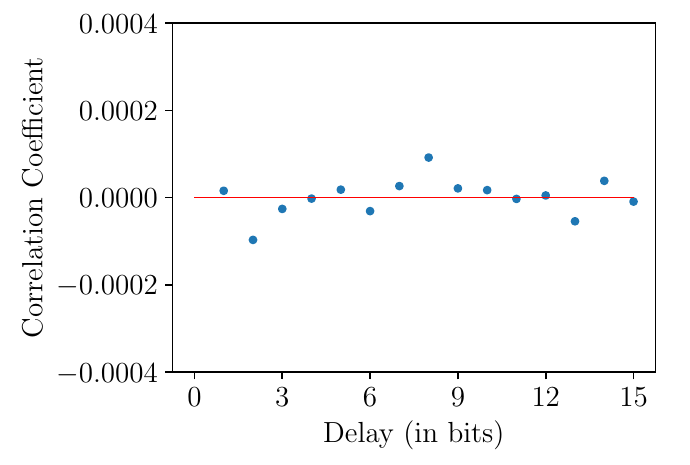}
        \caption{Region~5}
        \label{fig:pcoff5}
    \end{subfigure}
    \caption{Pearson correlation coefficients for $1$-bit to $15$-bit delays of the different $800\,\text{Mbit}$ samples generated from the arrival times of photons emitted by NV centres in the different regions of interest.}
    \label{fig:pcoffs}
\end{figure*}

\begin{table*}[]
\begin{tabular}{|l|l|l|l|l|l|l|l|}
\hline
\textbf{Statistical Quantity} & \textbf{Region 1} & \textbf{Region 2} & \textbf{Region 3} & \textbf{Region 4} & \textbf{Region 5} & \textbf{Ideal Value} \\ \hline
Entropy (per byte)             & 7.999998   & 7.999998   & 7.999998   & 7.999998   & 7.999998   & 8.000000   \\ \hline
$\chi^2$ Distribution          & 82.71\%    & 19.54\%    & 1.41\%     & 1.01\%     & 17.00\%    & 10--90\%   \\ \hline
Arithmetic Mean                & 127.492    & 127.507    & 127.484    & 127.496    & 127.492    & 127.500    \\ \hline
Monte Carlo value for $\pi$    & 3.14185045 & 3.14121709 & 3.14234461 & 3.14137861 & 3.14258869 & 3.14159265 \\ \hline
Serial Correlation Coefficient & 0.000375   & 0.000119   & 0.000145   & 0.000217   & 0.000658   & 0.000000   \\ \hline
\end{tabular}
\caption{ENT Statistical Test Suite results for the different $800\,\text{Mbit}$ samples generated from the arrival times of photons emitted by NV centres in the different regions of interest.}
\label{tab:ENT}
\end{table*}

In \figref{fig:multiple}, we plot both $p_{i}$ versus $i$ and $H_{\infty}$ versus $\lambda$ for the different regions of interest.  As the number of NV centres in a region increases, the non-uniformity of the probability distribution increases and the min-entropy decreases.  The decrease in min-entropy is marginal for small $\lambda$, but becomes substantial for larger $\lambda$.  The min-entropy values calculated for the different regions of interest are given in \tabref{tab:minent}, along with the experimentally measured values of $\lambda$.  Considering the precision of the values used to calculate the min-entropy, the min-entropy should be rounded to the ideal value of $1.00$ per bit for all the regions of interest.  This suggests that our experimental implementation of the time-of-arrival scheme with photons emitted by NV centres is capable of generating very high quality random numbers that are practically indistinguishable from uniformly-distributed true random numbers, for all the regions of interest.  As such, we do not anticipate a need for randomness extraction or any other form of post-processing.  In contrast, the min-entropy values estimated for implementations of the branching paths scheme for QRNG with photons emitted by NV centres are $0.00037$ per bit~\cite{QRNGbpNVC1} and $0.77$ per bit~\cite{QRNGbpNVC2}.  These previous implementations therefore needed intensive post-processing to improve the quality of the generated random numbers.

\subsection{Experimental results}\label{sec:QRNGexperiment}

The random number generation rates achieved for the different regions of interest are presented in \tabref{tab:rates}.  As expected, the generation rate increases as the number of NV centres in a region increases.  In particular, the generation rate ranges from $0.173\,\text{Mbits/s}$ for the single NV centre in Region~1 to $4.77\,\text{Mbits/s}$ for the cluster of just under 50 NV centres in Region~5.  In contrast, implementations of the branching paths scheme for QRNG with photons emitted by NV centres were only able to achieve generation rates of $34.37\,\text{bits/s}$~\cite{QRNGbpNVC1} and $0.76\,\text{Mbits/s}$~\cite{QRNGbpNVC2}.  Our implementation of the time-of-arrival scheme with photons emitted by the NV centres in Region~5 therefore demonstrates an order of magnitude improvement in generation rate compared to the highest generation rate previously achieved with NV centres.  In what follows, we apply some standard statistical tests to $800\,\text{Mbit}$ samples generated from the arrival times of photons emitted by NV centres in the different regions of interest.

\begin{table*}[]
\begin{tabular}{|l|l|l|l|l|l|}
\hline
\textbf{Work}                                 & \textbf{Speed (in Mbits/s)} & \textbf{NIST Tests} & \textbf{Randomness Extractor} & \textbf{Source} & \textbf{On-Chip} \\ \hline
Dynes \textit{et al.}~\cite{QRNGtoa2}         & 4.01                        & Passed              & None                          & Coherent        & No               \\ \hline
Nie \textit{et al.}~\cite{QRNGtoa5}           & 96                          & Passed              & Toeplitz Matrix Hashing       & Coherent        & No               \\ \hline
Khanmohammadi \textit{et al.}~\cite{QRNGtoa7} & 1                           & Passed              & XOR Hashing                   & Thermal (LED)   & Yes              \\ \hline
Strydom \textit{et al.}~\cite{QRNGtoa8}       & 14.4                        & Passed              & None                          & Coherent        & Waveguide        \\ \hline
Current Work (Region~1)                       & 0.173                       & Passed              & None                          & Single-photon   & Source           \\ \hline
Current Work (Region~5)                       & 4.77                        & Passed              & None                          & Multi-photon    & Source           \\ \hline
\end{tabular}
\caption{Comparison of our current implementation of the time-of-arrival scheme, with photons emitted by NV centres, with previously reported implementations of the time-of-arrival scheme.}
\label{tab:comparison}
\end{table*}

We first determine the relative frequency of $0$ and $1$ in the different $800\,\text{Mbit}$ samples.  The results are given in \tabref{tab:relfreq}.  The relative frequencies are around $0.5$ for all the regions of interest, which confirms that the bits generated by our experimental implementation of the time-of-arrival scheme are essentially uniformly distributed, for all the regions of interest, as predicted by the min-entropy estimation in \secref{sec:QRNGtheorymultiple}.  In \figref{fig:pcoffs}, we plot the Pearson correlation coefficient~\cite{Pearson} of each $800\,\text{Mbit}$ sample, obtained for each region of interest, with $1$-bit to $15$-bit delays of itself.  The Pearson correlation coefficient is a real number between $-1$ and $1$, where a positive value suggests that correlations are present, a negative value suggests that anti-correlations are present and a value close to $0$ suggests that no correlations are present.  It therefore follows that nearby bits are uncorrelated for all the regions of interest, as expected for high-quality random numbers.

For a more systematic assessment of the quality, we employ the industry standard ENT~\cite{ENT} and NIST~\cite{NIST} Statistical Test Suites.  In the ENT Statistical Test Suite, five important statistical quantities are calculated for each sample under investigation and the values obtained are compared to the ideal values for true random bits.  The ENT Statistical Test Suite results for the different regions of interest are presented in \tabref{tab:ENT}.  For all the regions of interest, the values obtained show excellent agreement with the ideal values for true random bits, as predicted by the min-entropy estimation in \secref{sec:QRNGtheorymultiple}.  The NIST Statistical Test Suite is specifically designed to assess a random number generator's suitability for use in cryptographic applications.  The NIST Statistical Test Suite passed for all five regions of interest, which verifies that the bits generated by our experimental implementation are of sufficient quality for use in cryptographic applications, for all five regions of interest.  This also confirms that our implementation does not require randomness extraction or any other form of post-processing.  A detailed list of comparisons with previous implementations of the time-of-arrival scheme is presented in \tabref{tab:comparison}.

\section{Conclusion}\label{sec:conclusion} 

We implemented the time-of-arrival scheme for QRNG with photons emitted by NV centres in fluorescent nanodiamonds.  To this end, we used a laser-scanning confocal microscopy setup to excite NV centres in fluorescent nanodiamonds and collect the emitted photons.  The second-order temporal correlation function was employed to determine the number of NV centres present in a nanodiamond or cluster of nanodiamonds.  In our experimental implementation of the time-of-arrival scheme, we investigated five regions of interest, namely: a nanodiamond with a single NV centre, a nanodiamond with two NV centres, a nanodiamond with four NV centres, a cluster of nanodiamonds with around 17 NV centres and a cluster of nanodiamonds with just under 50 NV centres.  The random number generation rates achieved range from $0.173\,\text{Mbits/s}$ for the single NV centre to $4.77\,\text{Mbits/s}$ for the cluster with just under 50 NV centres.  Hence, by employing multiple NV centres, we were able to increase the generation rate by an order of magnitude.  Furthermore, the generation rate achieved for the cluster with just under 50 NV centres also demonstrates an order of magnitude improvement compared to the highest generation rate previously achieved in an experiment investigating QRNG with NV centres~\cite{QRNGbpNVC2}.

To theoretically analyse the quality of the random numbers generated by our implementation, we employed the min-entropy.  In particular, we derived an expression for the min-entropy of random numbers generated from the arrival times of photons emitted by a three-level single-photon emitter~\cite{NVClevels} with background contamination, and generalised the result to $N$ statistically independent and identical single-photon emitters.  We applied this general theory to our experimental implementation and found that the min-entropy is approximately equal to the ideal value of one per bit for all five regions of interest.  This is in excellent agreement with our experimental results, since the bits generated by our experimental implementation passed the rigorous ENT~\cite{ENT} and NIST~\cite{NIST} Statistical Test Suites without any form of post-processing, for all five regions of interest.  Although our experiments were carried out with NV centres, both the experimental and theoretical results extend naturally to other defect centres in diamond and many other single-photon sources.  Furthermore, while our current implementation relies on an off-chip driving field to excite NV centres and the collection and detection of emitted photons is also performed off-chip, we note that future work on the integration of an on-chip electrical excitation system~\cite{NVCexcitation1, NVCexcitation2} and on-chip collection and detection~\cite{NVCdetection1, NVCdetection2} would enable a highly compact, highly robust, moderate speed QRNG chip.

\begin{acknowledgements}
We thank Jason Francis, Alex Huck and Andr\'{e} Smith for their assistance with the design and/or construction of the optical experimental setup for probing NV centre emitters.  This research was funded by the Harry Crossley Foundation and the South African Quantum Technology Initiative (SA~QuTI) through the Department of Science, Technology and Innovation (DSTI) of South Africa.
\end{acknowledgements}


\appendix

\onecolumngrid

\newpage

\section{Description of the confocal microscopy part of the optical setup}\label{append:confocal} 

The AL arrangement in \figref{fig:setup} is employed to collect diverging light beams from the nanodiamond sample, which in turn enables light from the sample to be captured.  The AL which is closest to the vertical microscope is one focal length from the rear aperture of the DLM objective, and bends the diverging beams from the sample so that they are roughly parallel to each other, thereby forming an image plane one focal length away, on the other side of the AL.  Another AL, placed one focal length from this image plane, is used to collimate and focus these parallel beams so that they converge to a point on the 2D GM (Thorlabs GVS202), again one focal length away.  After the 2D GM the beams from the sample, which contain fluorescence from the nanodiamonds as well as scattered light from the pump beam, diverge again.  When these diverging beams reach the DM that was used to direct the pump beam into the microscopy part of the setup, only fluorescence from the nanodiamonds, which consists mainly of wavelengths above $567\,\text{nm}$~\cite{NVCreview1, NVClevels, NVCentanglement}, is transmitted through the DM, while scattered light from the pump beam is reflected by the DM and is thus largely removed from the collection.  Another AL, placed one focal length from the point of convergence on the 2D GM, is then used to bend the diverging beams so that they become parallel again, forming a second image plane one focal length away, at a $50\,\mu\text{m}$ PH.  The two ALs which are closest to the 2D GM convert the angular movements of the 2D GM system into lateral translations of the image across the PH.  As such, the PH has conjugate points at the first image plane as well as on the sample.  Hence, for a given orientation of the 2D GM system, only light propagating from a specific conjugate point on the sample will pass though the PH, and any light from neighbouring points on the sample, as well as light from different depths will be blocked.

\section{Fluorescence spectra of the regions of interest}\label{append:spectra}

\begin{figure}[b]
    \centering
    \begin{subfigure}[b]{.329\textwidth}
        \centering
        \includegraphics[scale=0.5]{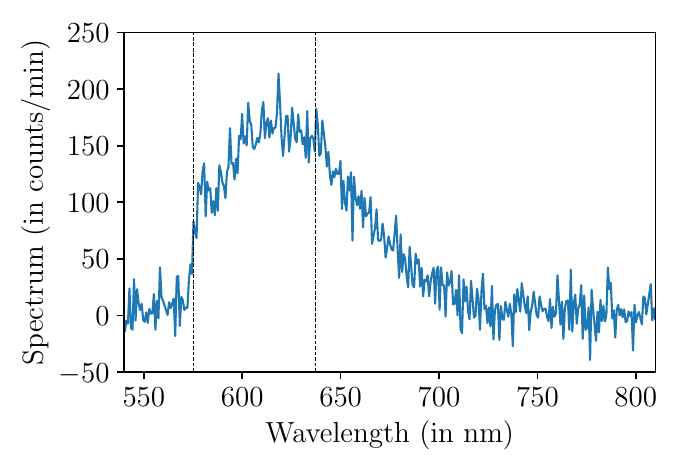}
        \caption{Region~1}
        \label{fig:spectrum1}
    \end{subfigure}
    \begin{subfigure}[b]{.329\textwidth}
        \centering
        \includegraphics[scale=0.5]{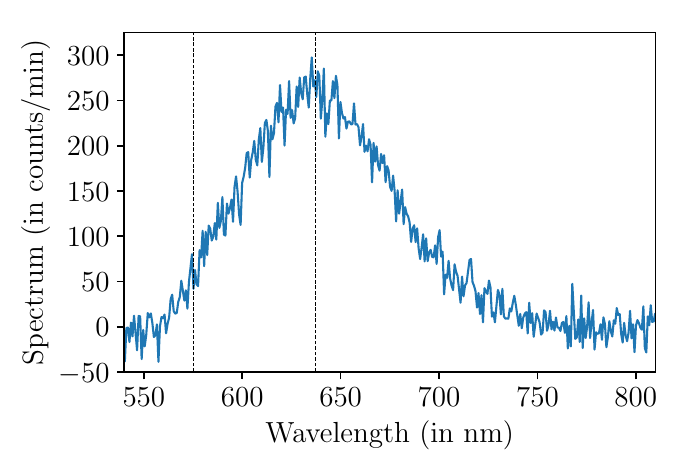}
        \caption{Region~2}
        \label{fig:spectrum2}
    \end{subfigure}
    \begin{subfigure}[b]{.329\textwidth}
        \centering
        \includegraphics[scale=0.5]{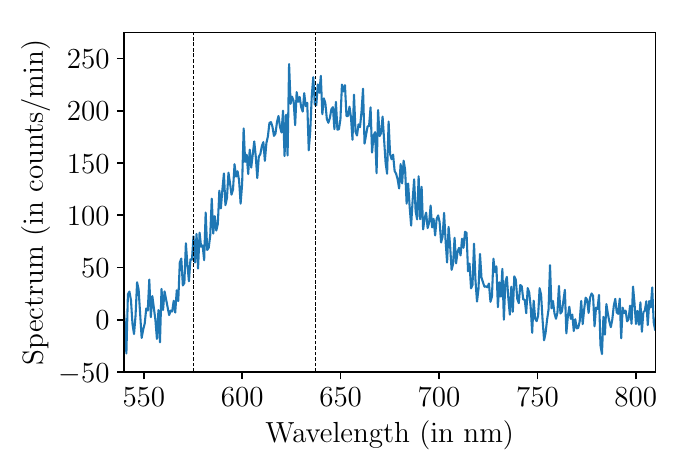}
        \caption{Region~3}
        \label{fig:spectrum3}
    \end{subfigure}

    \vspace{0.25cm}
    
    \begin{subfigure}[b]{.329\textwidth}
        \centering
        \includegraphics[scale=0.5]{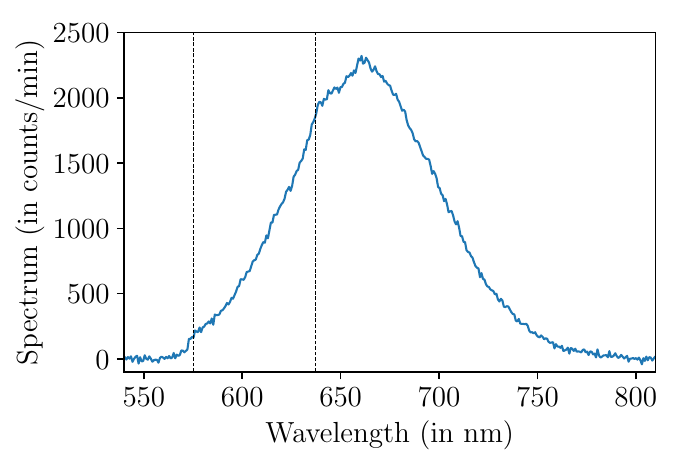}
        \caption{Region~4}
        \label{fig:spectrum4}
    \end{subfigure}
    \begin{subfigure}[b]{.329\textwidth}
        \centering
        \includegraphics[scale=0.5]{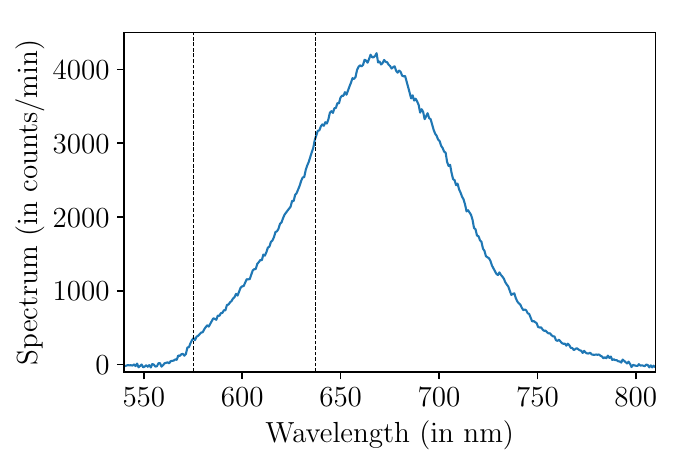}
        \caption{Region~5}
        \label{fig:spectrum5}
    \end{subfigure}
    \caption{Fluorescence spectra for the different regions of interest.  These fluorescence spectra are an average of 10 measurements, each with an integration time of $60\,\text{s}$.  The dashed black lines indicate the positions of the zero phonon lines of $\text{NV}^0$ and $\text{NV}^-$ centres.}
    \label{fig:spectra}
\end{figure}

To investigate the presence of $\text{NV}^0$ and $\text{NV}^-$ centres in the regions of interest, we measure their fluorescence spectra.  Both $\text{NV}^0$ and $\text{NV}^-$ centres have a broad fluorescence spectrum at room temperature.  In particular, $\text{NV}^0$ centres have a zero phonon line at $575\,\text{nm}$ with a phonon sideband that extends to $650\,\text{nm}$, while $\text{NV}^-$ centres have a zero phonon line at $637\,\text{nm}$ with a phonon sideband that extends to $800\,\text{nm}$~\cite{NVCreview1}.  To measure the fluorescence spectrum, we employ the configuration in \figref{fig:spectrumconfig}, that is we connect the SMF from the optical setup in \figref{fig:setup} to a spectrometer (Ocean Optics QE Pro-FL).  Furthermore, we extend the passband of the custom broadband BPF in \figref{fig:setup}, by removing the $600\,\text{nm}$ high-pass filter, so that the zero phonon line of the $\text{NV}^0$ will be visible.  The fluorescence spectra obtained for the different regions of interest are shown in \figref{fig:spectra}.  These suggest that a single $\text{NV}^0$ centre is present in Region~1 and that two $\text{NV}^0$ centres are present in Region~2, while both $\text{NV}^0$ and $\text{NV}^-$ centres are present in the other regions.

\section{Proof of uniformity for the time-of-arrival scheme for an arbitrary light source}\label{append:uniformity} 

Let $P_{t}(n)$ denote the probability of $n$ photons from an arbitrary light source arriving during a time interval of length $t$.  Consider a time interval of length $T$ divided into $M$ bins of equal width, $\tau=\frac{T}{M}$.  Under the assumption that at most one photon can arrive during time $T$, a photon arriving during this time arrives in an arbitrary bin $i$ with probability
\begin{equation}
p_{i} = \frac{(P_{\tau}(0))^{i-1}P_{\tau}(1)(P_{\tau}(0))^{M-i}}{\sum_{k=1}^{M}(P_{\tau}(0))^{k-1}P_{\tau}(1)(P_{\tau}(0))^{M-k}}.
\label{eq:uniformintermediate}
\end{equation}
The numerator is the probability that no photons arrive in the first $i-1$ bins, exactly one photon arrives in bin $i$ and no photons arrive in the remaining $M-i$ bins, which is precisely the probability that the one and only photon arriving during time $T$ arrives in bin $i$.  The denominator is the normalisation constant for the probabilities $p_{i}$ and is needed since we are assuming that exactly one photon arrives during time $T$.  The denominator is therefore the probability that one photon arrives during time $T$, which occurs whenever one photon arrives in any one of the $M$ bins and no photons arrive in all the other bins before and after it.  Simplifying we obtain,
\begin{equation}
p_{i} = \frac{P_{\tau}(1)(P_{\tau}(0))^{M-1}}{MP_{\tau}(1)(P_{\tau}(0))^{M-1}} = \frac{1}{M},
\label{eq:uniformfinal}
\end{equation}
and so a photon arrives in any one of the $M$ bins with a probability of $\frac{1}{M}$, so long as at most one photon can arrive during time $T$.  Since we did not make any assumptions about the form of $P_{t}(n)$, this result holds for any light source.

\section{Derivation of photon number probabilities for a single-photon emitter}\label{append:probabilities} 

To derive the photon number probabilities for a single-photon emitter, $P_{t}^{(1)}(n)$, we follow the approach of Ref.~\cite{NVCentanglement}.  The second-order temporal correlation function places an upper bound on the probability of multi-photon events during a time interval of length $t$~\cite{probabilities}.  In particular,
\begin{equation}
P_{t}^{(1)}(n \ge 2) \le \frac{1}{2}\mu_{t}^{2}g_{t}^{(2)}(0),
\label{eq:bound}
\end{equation}
where $\mu_{t}=\lambda t$ is the mean number of photons arriving during a time interval of length $t$, with $\lambda$ the time-independent mean photon flux, and $g_{t}^{(2)}(0)$ is the detected value of the second-order temporal correlation function, not simply at zero, but measured over a finite time interval of length $t$ around zero.  It can be calculated theoretically using
\begin{equation}
g_{t}^{(2)}(0) = \frac{2}{t^2}\int_{t'}^{t'+t}\int_{0}^{t'-t''+t}g^{(2)}(t''')\,dt'''\,dt''.
\label{eq:g2detectedformula}
\end{equation}
Substituting in the second-order temporal correlation function for a three-level single-photon emitter with background contamination (see \equatref{eq:g2background} with $N=1$) and integrating, we obtain
\begin{equation}
g_{t}^{(2)}(0) = 1-\frac{2\rho^{2}}{\gamma_{1}^{2}\gamma_{2}^{2}t^{2}}\left(\beta\gamma_{2}^{2}\left(e^{-\gamma_{1}t}+\gamma_{1}t-1\right)-(\beta-1)\gamma_{1}^{2}\left(e^{-\gamma_{2}t}+\gamma_{2}t-1\right)\right).
\label{eq:g2detectedresult}
\end{equation}

Since $\gamma_{1}$ is the rate associated with transitions between the ground state and the main excited state of the emitter, $\gamma_{1}$ is also the photon emission rate, and so for $t \ll \frac{1}{\gamma_{1}}$, we expect that $P_{t}^{(1)}(1) \gg P_{t}^{(1)}(2) \gg P_{t}^{(1)}(n \ge 3)$.  We therefore set $P_{t}^{(1)}(n) = 0$ for $n \ge 3$.  We also set $P_{t}^{(1)}(2)$ equal to its upper bound, that is
\begin{equation}
P_{t}^{(1)}(2) = \frac{1}{2}\mu_{t}^{2}g_{t}^{(2)}(0),
\label{eq:P2}
\end{equation}
to obtain the most conservative estimate of $P_{t}^{(1)}(2)$ within this approximation.  The mean photon number can be expressed as
\begin{equation}
\mu_{t} = \sum_{n=0}^{\infty}nP_{t}^{(1)}(n) = P_{t}^{(1)}(1) + 2P_{t}^{(1)}(2) + 3P_{t}^{(1)}(3) + \cdots,
\label{eq:mean}
\end{equation}
and normalisation ensures that
\begin{equation}
1 = \sum_{n=0}^{\infty}P_{t}^{(1)}(n) = P_{t}^{(1)}(0) + P_{t}^{(1)}(1) + P_{t}^{(1)}(2) + P_{t}^{(1)}(3) + \cdots,
\label{eq:norm}
\end{equation}
and so it follows that
\begin{equation}
P_{t}^{(1)}(1) = \mu_{t} - 2P_{t}^{(1)}(2) = \mu_{t} - \mu_{t}^{2}g_{t}^{(2)}(0),
\label{eq:P1}
\end{equation}
and that
\begin{equation}
P_{t}^{(1)}(0) = 1 - P_{t}^{(1)}(1) - P_{t}^{(1)}(2) = 1 - \mu_{t} + \frac{1}{2}\mu_{t}^{2}g_{t}^{(2)}(0).
\label{eq:P0}
\end{equation}

\end{document}